\documentclass[%
 aip,
 amsmath,amssymb,
 preprint,
]{revtex4-1}

\usepackage{graphicx}% Include figure files
\usepackage{dcolumn}% Align table columns on decimal point
\usepackage{bm}% bold math
\usepackage[utf8x]{inputenc}
\usepackage[T1]{fontenc}
\usepackage{mathptmx}
\usepackage{url}

\newcommand{\etal}{{\it et al.}}

\newcommand{\SpeciesFlux}{{F}}
\newcommand{\ReversibleStress}{{R}}
\newcommand{\WhiteNoiseMass}{{Z}}
\newcommand{\WhiteNoiseMomentum}{{W}}

\newcommand{\ViscousTensor}{\tau}
\newcommand{\MaxwellTensor}{\sigma}

\newcommand{\half}{\frac{1}{2}}
\newcommand{\deltachat}{\widehat{\delta c}}
\newcommand{\Dt}[1]{D_t^u #1}
\newcommand{\pderiv}[2]{\frac{\partial #1}{\partial #2}}

\usepackage{color}

\usepackage{float}
\usepackage[colorlinks=true,allcolors=blue]{hyperref}

\begin{document}

\title[Fluctuating hydrodynamics for ionic liquids]{A Low Mach Number Fluctuating Hydrodynamics Model For Ionic Liquids}
\author{Katherine Klymko}
\affiliation{Center for Computational Sciences and Engineering, Lawrence Berkeley National Laboratory}
\author{Sean Carney}
\affiliation{Department of Mathematics, The University of Texas at Austin}
\author{Andrew Nonaka}
\email{AJNonaka@lbl.gov}
\affiliation{Center for Computational Sciences and Engineering, Lawrence Berkeley National Laboratory}
\author{Alejandro Garcia}
\affiliation{Department of Physics and Astronomy, San Jose State University}
\author{John Bell}
\affiliation{Center for Computational Sciences and Engineering, Lawrence Berkeley National Laboratory}

\date{\today}

\begin{abstract}
We present a new mesoscale model for ionic liquids based on a low Mach number fluctuating hydrodynamics formulation for multicomponent charged species.
The low Mach number approach eliminates sound waves from the fully compressible equations leading to a computationally efficient incompressible formulation.
The model uses a Gibbs free energy functional that
includes enthalpy of mixing, interfacial energy, and electrostatic contributions.
These lead to a new fourth-order term in the mass equations and a reversible stress in the momentum equations.
We calibrate our model using parameters for [DMPI+][F6P-], an extensively-studied room temperature ionic liquid (RTIL), and numerically demonstrate the formation of mesoscopic structuring at equilibrium in two and three dimensions.
In simulations with electrode boundaries the measured double layer capacitance decreases with voltage, in agreement with theoretical predictions and experimental measurements for RTILs.
Finally, we present a shear electroosmosis example to demonstrate that the methodology can be used to model electrokinetic flows.
\end{abstract}

\maketitle

\section{Introduction}
An ionic liquid (IL) is a liquid salt with dissociated
cations and anions such as molten NaCl. 
Unlike conventional electrolyte solutions (e.g., seawater), an ionic liquid does not require a polar solvent.
Of particular interest are ionic liquids
composed of complex hydrocarbons that are high-viscosity liquids at room temperature.
These room temperature ionic liquids (RTILs) exhibit intriguing physical properties such as
high charge density \cite{SilvesterCompton_2009,Wishart_2009} and extremely low vapor pressures \cite{tokuda2006}.
Such properties make them attractive for energy technology applications such as super-capacitors \cite{brandtETAL_2013},
batteries \cite{LEWANDOWSKI_2009},
and
dye-sensitized photoelectrochemical cells \cite{LiETAL_2014}.
RTILs also have technological applications as designer solvents in areas such as
lubrication of micro-electromechanical machines \cite{lubricants_2009,lubricants_2013}.

Room temperature ionic liquids exhibit a number of interesting physical features. 
Capacitance measurements with RTILs show different behavior as a function of applied voltage than conventional electrolytes, which
reflects overcrowding at the electrode surfaces resulting from the large size of the ions~\cite{DoubleLayerCapacitance}. 
At low voltages, short-range Coulomb interactions also lead to overscreening in which the layer next to an
electrode has excess charge relative to the electrode, resulting in 
the formation of a subsequent, weaker layer of opposite charge~\cite{bazant2011double}.
Molecular dynamics simulations \cite{MerletETAL_2014,maginn2007atomistic,rajput2012structure}  and experimental evidence \cite{PanFreyland_2006,SuETAL_2009,WenETAL_2015} show that RTILs are heterogeneous at nanoscale levels, forming phase separation of
anions and cations on scales of a few nanometers. 

%\Klymko{say a bit more here--need to go through Limmer argument a bit more carefully; phase transition in the ordering of the anions and cations? Or potentially something only applicable in three-dimensional systems. We need to distinguish our phase transition from the spike in capacitance Limmer discusses. also, ours is a second order transition, right? Or maybe best not to say anything at all}.

Strong inter-ionic correlations and structure render classical models such as Nersnt-Planck
%\MarginPar{The Ginzburg-Landau type of free energy that we're using is something I'd call a "classical mean field theory" so I'm not sure what this sentence means. AG Is this statement pulled from some other RTIL paper?  have i address this? JBB Yes, it's good now and the paper this seems to come from said "existing classical mean field theories". AG} 
used to describe dilute electrolytes inapplicable. Kornyshev and co-workers \cite{DoubleLayerCapacitance,GoodwinETAL_2017} address the impact of ion size and enthalpy of mixing on the structure of the electrical double layer (EDL) in an ionic liquid. Their model gives a diffuse double-layer capacitance that extends the classical Guoy-Chapman theory and is in agreement with experimental measurements~\cite{jitvisate2018direct}.
Bazant, Story, and Kornyshev \cite{BSK_2011} develop a Landau-Ginzburg-like model that includes effects of ion size and
overscreening that is able to predict the structure of the EDL, and improves the prediction of
the capacitance.
Limmer \cite{Limmer_2015} introduces a mean-field model that incorporates short range repulsion between cations and anions.
The interplay of this short-range repulsion with electrostatic forces then determines the morphology of the
ionic liquid. 
Gavish and Yochelis \cite{GavishYochelis_2016} construct
a model by adding a Flory-Huggins-like term
and an electrostatic term to the free energy of an ideal fluid. 
 The resulting system is similar to the Ohta-Kawasaki
model \cite{OhtaKawasaki_1986,OhtaKawasaki_1988} for diblock co-polymers coupled to an electric field.  They 
solve the resulting diffusion equation coupled to the electrostatic equation to study structure in the
bulk and how the structure couples to the electric double layer.  

Simulation models for ionic liquids generally fall into two categories: coarse-grained lattice models~\cite{Limmer_2015, DoubleLayerCapacitance,bozym2015anomalous} and detailed microscopic models such as molecular dynamics~\cite{MerletETAL_2014,maginn2007atomistic} and density functional theory~\cite{jiang2011density,katsyuba2007application}. The former have the advantage of capturing qualitative features of an ionic liquid while being computationally efficient. They have the disadvantage of omitting many physical details required for quantitative predictions. On the other hand, microscopic models capture molecular detail but they are computationally demanding and additionally require delicate tuning of the potentials for the complex molecules in an ionic liquid.

This paper introduces a new mesoscopic simulation model for ionic liquids. 
Specifically, we develop a low Mach number fluctuating hydrodynamics model that is similar to the work of Lazaridis et al. \cite{lazaridis2017}.
Their model is based on a compressible isothermal formulation that has a more comprehensive description of the
fluid than in Gavish and Yochelis \cite{GavishYochelis_2016}, albeit with a somewhat idealized thermodynamic representation.
The deterministic component of the model here is similar to the quasi-incompressible Cahn-Hilliard fluid model of Lowengrub and 
Truskinovsky \cite{lowengrub_1998} coupled to an electric field.  The incorporation of stochastic terms allows our mesoscopic model to capture the effects of thermal fluctuations which cannot be neglected at the nanometer scale, the length scale at which typical structures
form in ionic liquids. The low Mach number formulation analytically removes sound waves from the model equations based on the assumption that they do not significantly affect the system dynamics. 
This eliminates the acoustic time step restriction allowing for time steps that are two or more orders of magnitude larger than the comparable compressible formulation, and significantly more computationally efficient than MD simulations of a similar size.

%\Klymko{timestep for MD is actually similar to what is needed for compressible equations (we're using $2*10^-13$, think Voulgarikis is doing $10^-15$ and MD would also be using femtoseconds)}

The outline of the paper is as follows: first, the Gibbs free energy functional is defined and the fluctuating hydrodynamic equations of motion are outlined in Section \ref{sec:formulation}. In particular the free energy contains ``excess'' and nonlocal contributions that model repulsive forces between cation and anion and interfacial tension, respectively. These contributions are calibrated to roughly match the feature size of a typical RTIL as %, and their specific form is 
determined from a stability analysis of the concentration equation. After a description of the numerical methods used to discretize the equations of motion in Section \ref{sec:numerical_methods}, numerical results are presented in Section \ref{sec:simulation_results}.
First we show the bulk morphology in both two and three dimensions. Then we discuss the dependence of capacitance on voltage, comparing with the theoretical predictions of Goodwin \etal \cite{GoodwinETAL_2017}.
We show that the structure of the electric double layer at the electrodes changes significantly if thermal fluctuations are omitted, as previously observed by Lazaridis \etal\cite{lazaridis2017}. Finally, we demonstrate the capability of modeling electrokinetic flows with a simulation of electroosmotic shear. Section \ref{sec:conclusions} concludes with a discussion of the results and their implications for future work.%\MarginPar{rewrote a bit JBB. Looks good to me. AG}

%In our work, we define the excess free energy and the coefficient of the non-local term in the free energy functional
%to roughly match the feature size of a typical RTIL.  Numerical results are presented demonstrating the
%bulk morphology in both two and three dimensions, the dependence of capacitance on voltage, and the structure of the
%electric double layer in the presence of electrodes at the boundary. We compare our capacitance results with theoretical predictions.

\section{Formulation}\label{sec:formulation}

Our goal here is to develop a low Mach number model for room temperature ionic liquids.  
We introduce a free energy
functional similar to Gavish and Yochelis \cite{GavishYochelis_2016} that includes enthalpy of mixing, interfacial energy, and electrostatic contributions.
%\st{adds an excess free energy and
%charge effects to an ideal fluid.}
%\Klymko{I think I'm okay with this, but we should make sure we all agree--I'll need to change notation further down ($g^{ex}$ for example; we could go with $g^{enth}$ but that would presumably also include the ideal part, right?)}\Garcia{We don't have to change notation. Gibbs free energy is $G = H - TS$ where $H$ is enthalpy. The concentration dependence of $G$ is expressed as being in two parts, enthalpy and entropy, in the thermodynamic theories for solutions, such as Flory-Huggins. But I prefer $g^{ex}$ to $g^{enth}$ for the notation.}.\MarginPar{Is this okay or need further discussion? AG} 
Based on that free energy functional we then develop a low Mach number fluctuating hydrodynamics model for
ionic liquids by extending the methodology developed in a series of papers
\cite{donev2014low,nonaka2015low,donev2015low,peraud2016,donev2019fluctuating} for multispecies mixtures of charged ionic fluids.  
For simplicity, we adopt an isothermal two-species approximation and assume that the two species, the
cation and the anion, have the same molecular mass $m$ and equal but opposite charge.
We assume each species is incompressible and has the same density; hence, the velocity field satisfies an incompressibility constraint.

We write the Gibbs free energy as
\begin{equation}
G\left(c,\nabla c, \phi,\nabla \phi\right) = \int \left[\rho \; g(c, \nabla c) + \rho z (2c-1) \phi - \frac{1}{2} \epsilon |\nabla \phi|^2 \right] d\mathbf{r}
\label{eq:FE}
\end{equation}
where $c$ denotes the cation concentration, $\phi$ is the electric potential, $\epsilon$ is the (constant) static permittivity, $\rho$ is the density
and $z$ is the charge per mass of cation.  The non-electrical contribution to the specific free energy is given by
\begin{equation}\label{eq:g_no_electrostatic}
g(c,\nabla c ) = \frac{k_B T}{m} \left[ g^{\rm id}(c) + g^{\rm ex}(c) + \frac{\kappa}{2} |\nabla c|^2 \right ]
\end{equation}
where $T$ is temperature, $k_B$ is Boltzmann's constant, and $\kappa$ is an interfacial parameter, which is assumed to be constant.
The entropy of mixing contribution is
\begin{equation}
g^{\rm id}(c) = c \; \log (c) + (1-c) \; \log (1-c) 
\end{equation}
and $g^{\rm ex}$ is an excess free energy due to the enthalpy of mixing.
Note that Lazaridis et al. \cite{lazaridis2017} include a contribution to the free energy $G$ that depends on $\nabla \rho$; however, the resulting term will vanish in the low Mach number flow limit %\MarginPar{more appropriate to refer to the low-mach number limit instead? }.

For systems in which the characteristic fluid velocity is asymptotically small relative to the sound speed, we can obtain the low Mach number equations from the fully compressible equations by asymptotic analysis \cite{Klainerman:1982,Majda:1985}. Taking density, $\rho_0$, as constant the equations of motion are
\begin{eqnarray}
( \rho_0 c)_t + \nabla \cdot(\rho_0 u c) &=& \nabla \cdot {\SpeciesFlux}   \nonumber \\
( \rho_0 u)_t + \nabla \cdot(\rho_0 u u) + \nabla \pi &=& \nabla \cdot {\ViscousTensor}  + \nabla \cdot \MaxwellTensor  +
\nabla \cdot \ReversibleStress \nonumber \\
\nabla \cdot u &=& 0 \nonumber \\
-\epsilon \nabla^2 \phi &=&  q^f \equiv \rho_0 z (2c-1),
\label{eq:low_mach_eqs}
\end{eqnarray}
where $u$ is the fluid velocity, $\pi$ is a perturbational pressure, and $q^f$ is the charge density.
Here, $\SpeciesFlux$, $\ViscousTensor$, $\MaxwellTensor$ and $\ReversibleStress$ are the species flux, viscous stress tensor,
Maxwell stress tensor, and the interfacial reversible stress, respectively.

%%%%%%%%%%%%%%%%%%% Compressible %%%%%%%%%%%%%%%%%%%%
% With the above assumptions, the isothermal, compressible equations of motion are given by\Klymko{do we want to put a bit more of a derivation--maybe in an appendix?}
% \begin{eqnarray}
% ( \rho c)_t + \nabla \cdot(\rho u c) &=& \nabla \cdot {\SpeciesFlux}   \nonumber \\
% ( \rho u)_t + \nabla \cdot(\rho u u) + \nabla p &=& \nabla \cdot {\ViscousTensor}  + \nabla \cdot \MaxwellTensor  +
% \nabla \cdot \ReversibleStress\\
% \rho_t + \nabla \cdot (\rho u) &=& 0 \nonumber \\
% -\epsilon \nabla^2 \phi &=& q^f \equiv \rho z (2c-1) \nonumber
% \label{eq:comp_eqs}
% \end{eqnarray}
% where $u$ is the fluid velocity, $p$ is the pressure, and $q^f$ is the charge density.
% Here, $\SpeciesFlux$, $\ViscousTensor$, $\MaxwellTensor$ and $\ReversibleStress$ are the species flux, viscosity tensor,
% Maxwell stress tensor, and the \GarciaV{interfacial reversible stress \st{reversible stress due to interfacial tension}}, respectively.
%%%%%%%%%%%%%%%%%%%%%%%%%%%%%%%%%%%%%%%%%%%%%%%%%%%%%%%%%%%%%%%

In the fluctuating hydrodynamics model, the dissipative fluxes, $\SpeciesFlux$ and $\ViscousTensor$, contain both deterministic and stochastic terms, e.g., $\SpeciesFlux = \overline{\SpeciesFlux} + \widetilde{\SpeciesFlux}$.
The deterministic species flux can be represented in Onsager form as \cite{DM_63}
\begin{equation}\label{eq:spec_flux}
\overline{\SpeciesFlux} = L \frac{\nabla_T \mu}{T} 
\end{equation}
where $\mu$ is the difference in electro-chemical potential between cations and anions, $\nabla_T$ refers to the gradient with $T$ held fixed, and $L$ is an Onsager coefficient. 
Differentiation of the specific free energy with respect to $c$ yields 
%\MarginPar{Why is this a partial derivative instead of just $d/dc$~? AG No good reason. think forward to additional dependencies -- JBB Okay, just leave it as it is then. AG}
\begin{equation}
\mu(c) = \frac{k_B T}{m} \left [   
\log (c) - \; \log(1-c) + \frac{\partial  g^{ex}(c)}{\partial c} + \kappa \nabla^2 c
\right ]
+ 2 z \phi.
\label{eq:mu}
\end{equation}
%
%Note that since $\mu = \mu_1 - \mu_2$, for each species
%\begin{equation}
%%\log (x_i)  + \frac{\partial  g^{ex}(x_i)}{\partial x_i} + \frac12 \kappa \nabla^2 x_i
%\right ]
%+ z_i \phi
%\end{equation}
%\MarginPar{keep the$\mu_i$??}
%where $x_1 = c$, $x_2 = 1-c$, and $z = z_1 = - z_2$. 
%
Inserting (\ref{eq:mu}) into \eqref{eq:spec_flux} then gives
\begin{equation}
\overline{\SpeciesFlux} = L \left [ \frac{k_B }{m} \left( \frac{1}{c}+\frac{1}{1-c} + 
\frac{\partial^2 g^{ex}(c)}{\partial c^2} \right) \nabla c +\frac{k_B \kappa }{m} \nabla \nabla^2 c + 2 z \nabla \phi \right ] .
\end{equation}
For a two component mixture, the Onsager coefficient is given by \cite{Landau_59}
\begin{equation}
L = \frac{\rho_0 m }{k_B} D c (1-c)
\end{equation}
where $D$ is the Fickian binary diffusion coefficient.
 The species flux expressed in terms of $D$ then is
\begin{equation}
\label{deterministic_flux}
\overline{\SpeciesFlux} = \rho_0 D \left ( \nabla c +  c (1-c)
\frac{\partial^2 g^{ex}(c)}{\partial c^2} \nabla c +  c(1-c) \kappa \nabla \nabla^2 c +
\frac{2 z m c (1-c)}{k_B T} \nabla \phi  \right ).
\end{equation}
The amplitude of the noise satisfies a
fluctuation dissipation relation \cite{Zarate_07, Kubo_1966}
% \st{The coefficient of the noise term is given by $\sqrt{2 k_B L}$ \cite{Zarate_07}
% so that} 
\begin{equation}  %\label{eq:species_noise}
\widetilde{\SpeciesFlux} = \sqrt{2 k_B L} ~\WhiteNoiseMass = \sqrt{2 \rho_0 m D c (1-c)}  ~\WhiteNoiseMass
\label{stochastic_flux}
\end{equation}
where $\WhiteNoiseMass(\mathbf{r},t)$ is standard, uncorrelated Gaussian white noise.

The viscous stress tensor is given by
$\ViscousTensor =  \overline{\ViscousTensor} +  \widetilde{\ViscousTensor}$
where the deterministic component
\begin{equation}
\overline{\ViscousTensor} = \eta [\nabla u + (\nabla u)^T]
\end{equation}
and $\eta$ is viscosity.  Here, bulk viscosity is neglected because it does not appear in the low Mach number equations.
The stochastic contribution to the viscous stress tensor is modeled as,
\begin{equation}
\widetilde\ViscousTensor = \sqrt{\eta k_B T}({\WhiteNoiseMomentum} + {\WhiteNoiseMomentum}^T),
\end{equation}
where 
${\WhiteNoiseMomentum}(\mathbf{r},t)$ is a standard Gaussian white noise tensor
with uncorrelated components,
\begin{equation}
\langle {\WhiteNoiseMomentum}_{ij}(\mathbf{r},t)
{\WhiteNoiseMomentum}_{kl}(\mathbf{r}',t') \rangle =
\delta_{ik}\delta_{jl} \delta(\mathbf{r}-\mathbf{r}') \delta(t-t'),
\end{equation}
and, again, the amplitude of the noise satisfies a
fluctuation dissipation relation \cite{Zarate_07, Kubo_1966}. 

To complete the specification of the model we need to define $\MaxwellTensor$ and $\ReversibleStress$. In the absence of a magnetic field \cite{landau1992electrodynamics}, the Maxwell stress is 
\begin{equation}
\MaxwellTensor_{ij} = \epsilon E_i E_j - \half \epsilon E^2 \delta_{ij},
\end{equation}
where $E = -\nabla \phi$.
Assuming a constant static permittivity,
$\epsilon \nabla\cdot {E} = q^f$, so 
the resulting force density on the fluid is,
\begin{equation}\label{eq:lorentz_force}
\nabla\cdot \MaxwellTensor = q^f {E} = - q^f \nabla \phi,
\end{equation}
which is simply the Lorentz force.
The interfacial reversible stress 
\begin{equation}
\ReversibleStress = \frac{\rho_0 k_B T \kappa}{m} \left [\frac{1}{2} |\nabla c|^2 \mathbb{I} - \nabla c \otimes \nabla c \right ]
\label{reversible_stress}
\end{equation}
is derived from a variational principle as detailed in Appendix \ref{section:EL_appendix};
see also \citenum{lowengrub_1998,shang2011fluctuating,lazaridis2017}.
Note that since both $\MaxwellTensor$ and $\ReversibleStress$ are non-dissipative fluxes, they have no corresponding stochastic fluxes.

For boundary conditions, in this paper we consider two types: periodic boundaries and no-slip impermeable electrode walls.
In the latter case, the velocity at the wall is zero and the electric potential satisfies a Dirichlet condition.
For concentration, we specify that both the normal derivative and the total flux vanish at walls.  Spatial discretization details for these boundary conditions are described in Section \ref{sec:spatial}.

\section{Numerical method}\label{sec:numerical_methods}
The equations of motion \eqref{eq:low_mach_eqs} consist of species transport and momentum evolution with an incompressibility constraint on the velocity field coupled to a Poisson equation for the electric potential.
The system is discretized in a structured-grid finite-volume approach with cell-averaged concentrations and face-averaged (staggered) velocities.  Integration in time is performed with a predictor-corrector scheme. 
Below we summarize our spatial and temporal discretization, noting that we are building off the explicit electrodiffusion approach used in Donev \etal \cite{donev2019fluctuating}, except here we do not consider reactions.
Here, the two primary additions are the inclusion of the excess free energy and interfacial terms in the deterministic mass flux \eqref{deterministic_flux} and the reversible stress tensor in the momentum equation \eqref{reversible_stress}.
 
\subsection{Spatial Discretization}\label{sec:spatial}
As detailed in~\citenum{donev2014low, Donev_10}, the spatial discretizations of the equations for mass and momentum transport are based on standard second-order stencils for derivatives and spatial averaging to ensure a discrete fluctuation-dissipation balance.
The electrodiffusion term in the species fluxes and the Lorentz force in the momentum equation are computed from the electric potential.
This potential is obtained by solving Poisson's equation with a cell-centered multigrid solver~\cite{peraud2016}.
The multigrid solver uses standard second-order stencils, and supports user-specified Dirichlet conditions on the potential for electrode wall boundary conditions.
For velocity we set the velocity field to zero on walls and use
one-sided approximations to evaluate the viscous stress.
%to zero, set the evaluation of the viscous, advective, and stochastic terms on the wall to zero, and linearly extrapolate tangential velocity to ghost cells to retain the use of the same stencils everywhere.
The random numbers for the stochastic contribution to the viscous stress tensor are generated on shifted control volumes about each cell face.
We note that for tangential velocities adjacent to no-slip walls, there is a stochastic flux on the wall itself; this noise term has twice the variance of the noise in the bulk \cite{Balboa2012}.

Since the interfacial tension term in \eqref{eq:g_no_electrostatic} introduces a new, third-order term in the species flux $\SpeciesFlux$ and a reversible stress tensor in the momentum equation, their discretization is described in detail here. 
The center of the cells in two dimensions are indexed by $(i,j)$ and the faces along $x,y$ as $(i+\half,j),(i,j+\half)$, where $i=1,...,N_x, j=1,...,N_y$. 
The species diffusion fluxes are computed on the faces of the grid based on \eqref{deterministic_flux} and \eqref{stochastic_flux}, and the divergence of the flux is approximated with
\begin{equation}
    \left( \nabla \cdot \SpeciesFlux \right)_{i,j} \approx \Delta x^{-1} \left[ \SpeciesFlux^{(x)}_{i+\half,j} - \SpeciesFlux^{(x)}_{i-\half,j}\right] + \Delta y^{-1}\left[ \SpeciesFlux^{(y)}_{i,j+\half} - \SpeciesFlux^{(y)}_{i,j-\half} \right] \ .
\end{equation}
The new third order term in the species flux equation  \eqref{deterministic_flux} (i.e., the term proportional to $ \kappa \nabla \nabla^2 c$) is computed by first approximating $\nabla^2 c$ at cell centers. Here, nine and twenty-one point stencils in two and three dimensions, respectively, are used so that the discrete Laplacian is more isotropic numerically and hence reflective of the isotropic contribution $\kappa/2 |\nabla c|^2$ to the free energy density.
Specifically, if the undivided difference operator in two dimensions is defined as
\[
\delta_x^2 c_{i,j} = c_{i+1,j} -2 c_{i,j}+ c_{i-1,j}
\]
(with $\delta_y^2$ defined analogously), then the Laplacian is approximated by
\begin{equation}
\nabla^2 c_{i,j} \approx \left[\left(I + \frac{\delta_y^2}{6}\right) \frac{\delta_x^2}{\Delta x^2}+
\left(I + \frac{\delta_x^2}{6}\right) \frac{\delta_y^2}{\Delta y^2} \right ] c_{i,j}.
\end{equation}
%as follows for two dimensions where $c_{i,j}$ represents the average value of c on cell $i,j$:
%\begin{align}
%     \nabla^2 c_{i,j} &= \frac{1}{3 \Delta x^2} \big[ c_{i+1,j-1} + c_{i,j-1} + c_{i-1,j} + c_{i-1,j-1} \\ \nonumber
%     &+ c_{i+1,j} + c_{i,j+1} + c_{i+1,j+1} + c_{i-1,j+1} - 8c_{i,j} \big] \ ,
%\end{align}
The generalization to three dimensions is then:
\begin{eqnarray}
\nabla^2 c_{i,j,k} \approx \left[\left(I + \frac{\delta_y^2}{6}\right)
\left(I + \frac{\delta_z^2}{6}\right)\frac{\delta_x^2}{\Delta x^2}
+
\left(I + \frac{\delta_x^2}{6}\right)
\left(I + \frac{\delta_z^2}{6}\right)\frac{\delta_y^2}{\Delta y^2} 
+
\left(I + \frac{\delta_y^2}{6}\right)
\left(I + \frac{\delta_z^2}{6}\right)\frac{\delta_z^2}{\Delta z^2}
 \right ] c_{i,j,k}. \nonumber \\
%    \nabla^2 c_{i,j,k} &= \frac{1}{12 \Delta x^2} \big[ c_{i-1,j+1,k+1}+c_{i+1,j+1,k+1}+c_{i+1,j-1,k+1}+c_{i-1,j-1,k+1}  \nonumber \\ 
%    &+2c_{i-1,j,k+1}+2c_{i,j-1,k+1}+2c_{i,j+1,k+1}+2c_{i+1,j,k+1} \nonumber \\
%    &+c_{i-1,j+1,k}+c_{i+1,j+1,k}+c_{i+1,j-1,k}+c_{i-1,j-1,k}-32c_{i,j,k} \nonumber \\
 %   &c_{i-1,j+1,k-1}+c_{i+1,j+1,k-1}+c_{i+1,j-1,k-1}+c_{i-1,j-1,k-1}  \nonumber \\ 
%    &+2c_{i-1,j,k-1}+2c_{i,j-1,k-1}+2c_{i,j+1,k-1}+2c_{i+1,j,k-1}  \big]
\end{eqnarray}
Discrete gradients of the Laplacian are then computed at cell faces and added to the other terms in the deterministic species diffusion flux. For cells adjacent to the boundary, the evaluation of the Laplacian reflects the vanishing of the normal derivative of concentration.
At impermeable walls we also set the total species concentration fluxes to zero; i.e., the sum of deterministic flux and the stochastic mass fluxes on walls is set to  zero.

The other new term in the low Mach model is the reversible stress tensor~\eqref{reversible_stress} in the momentum equation.  The discretization here is somewhat more complex because of the use of staggered velocities; terms appearing in the $x$ velocity need to be evaluated a $x$ faces, etc. The first step is to compute the gradients of $c_{i,j}$ at grid nodes--in two
dimension these are
\begin{align}
 %   \nabla_x c_{i,j} &= \frac{1}{2\Delta x} \left[ c_{i,j} - c_{i-1,j}+c_{i,j-1}-c_{i-1,j-1} \right] \\
%    \nabla_y c_{i,j} &= \frac{1}{2\Delta y} \left[ c_{i,j} - c_{i,j-1}+c_{i-1,j}-c_{i-1,j-1} \right]    G^n_y c_{i,j} &= \frac{1}{2\Delta y} \left[ c_{i,j} - c_{i,j-1}+c_{i-1,j}-c_{i-1,j-1} \right]
  ( G^n_x c)_{i+\half,j+\half} &= \frac{1}{2\Delta x} \left( c_{i+1,j} - c_{i,j}+c_{i+1,j+1}-c_{i,j+1} \right) \\
  (G^n_y c)_{i+\half,j+\half} &= \frac{1}{2\Delta y} \left( c_{i,j+1} - c_{i,j}+c_{i+1,j}-c_{i+1,j} \right).  
\end{align}
The nodal gradients are then averaged to cell centers
\begin{align}
%    \overline{c_{x^+}} &= \frac{1}{4}\left[ \nabla_x c_{i,j} + \nabla_x c_{i,j+1}+\nabla_x c_{i+1,j+1}+\nabla_x c_{i+1,j}  \right] \\
%    \overline{c_{x^-}} &= \frac{1}{4}\left[ \nabla_x c_{i,j} + \nabla_x c_{i,j+1}+\nabla_x c_{i-1,j+1}+\nabla_x c_{i-1,j}  \right] \ .
    ( G^{cc}_x c)_{i,j} &= \frac{1}{4}\left[ (G^n_x c)_{i+\half,j+\half} + (G^n_x c)_{i-\half,j+\half}+(G^n_x c)_{i+\half,j-\half}+(G^n_x c)_{i-\half,j-\half}  \right] \\
   ( G^{cc}_y c)_{i,j} &= \frac{1}{4}\left[ (G^n_y c)_{i+\half,j+\half} + (G^n_y )c_{i-\half,j+\half}+(G^n_y c)_{i+\half,j-\half}+(G^n_y c)_{i-\half,j-\half}  \right] \ .
\end{align}
From this one can define a second order approximation to $\nabla\cdot \ReversibleStress$ by using conservative 
differences of the nodal and cell averaged gradients as: 
\begin{eqnarray}
 %   f_{x,i,j}&= - \kappa \frac{1}{\Delta y} \left( \nabla_x c_{i,j+1}\nabla_y c_{i,j+1} - \nabla_x c_{i,j}\nabla_y c_{i,j}  \right) \nonumber \\ 
%    & + \kappa \frac{1}{\Delta x}\left( \frac{1}{2}\left( \overline{c_{y^+}}^2-\overline{c_{x^+}}^2\right) - \frac{1}{2}\left( \overline{c_{y^-}}^2-\overline{c_{x^-}}^2 \right) \right) \\
%    f_{y,i,j}&= - \kappa \frac{1}{\Delta x} \left( \nabla_x c_{i+1,j}\nabla_y c_{i+1,j} - \nabla_x c_{i,j}\nabla_y c_{i,j}  \right) \\ \nonumber
 %   & + \kappa \frac{1}{\Delta y}\left( \frac{1}{2}\left( \overline{c_{y^+}}^2-\overline{c_{x^+}}^2\right) - \frac{1}{2}\left( \overline{c_{y^-}}^2-\overline{c_{x^-}}^2 \right) \right)      
     (\nabla \cdot \ReversibleStress )_{x,i+\half,j} &\approx& 
     - \frac{\rho_0 k_B T \kappa}{m} \frac{1}{\Delta y} \left[(G^n_x c)_{i+\half,j+\half}(G^n_y c)_{i+\half,j+\half} - (G^n_x c)_{i+\half,j-\half}(G^n_y c)_{i+\half,j-\half}  \right] \nonumber \\ 
    && +\frac{\rho_0 k_B T \kappa}{m} \frac{1}{\Delta x}\left[\frac{1}{2}\left( (G^{cc}_{y} c)_{i+1,j}^2-(G^{cc}_x c)_{i+1,j}^2\right) -
    \frac{1}{2}\left( (G^{cc}_{y}c)_{i,j}^2-(G^{cc}_{x}c)_{i,j}^2\right)  \right] \\
    (\nabla \cdot \ReversibleStress )_{y,i,j+\half}&\approx& 
    - \frac{\rho_0 k_B T \kappa}{m} \frac{1}{\Delta x} \left[(G^n_x c)_{i+\half,j+\half}(G^n_y c)_{i+\half,j+\half} - (G^n_x c)_{i-\half,j+\half}(G^n_y c)_{i-\half,j+\half}  \right] \nonumber \\ 
    && +\frac{\rho_0 k_B T \kappa}{m}  \frac{1}{\Delta y}\left[\frac{1}{2}\left( (G^{cc}_{y}c)_{i,j+1}^2-(G^{cc}_{x}c)_{i,j+1}^2\right) -
    \frac{1}{2}\left((G^{cc}_{y}c)_{i,j}^2 - (G^{cc}_{x}c)_{i,j}^2\right)  \right],
  %  & + \kappa \frac{1}{\Delta y}\left( \frac{1}{2}\left( \overline{c_{y^+}}^2-\overline{c_{x^+}}^2\right) - \frac{1}{2}\left( \overline{c_{y^-}}^2-\overline{c_{x^-}}^2 \right) \right)      
\end{eqnarray}
%where $f_x$ and $f_y$ are force densities in the $x$ and $y$ velocity equations, respectively.  
The reversible stress tensor in three dimensions is treated analogously.

\subsection{Temporal Discretization}

The basic temporal discretization is a predictor-corrector scheme for both concentration and velocity.
Given the values $c^n$ and $u^n$ at the beginning of time step $t^n$, the method consists of a preliminary step to obtain the concentration and velocity at $t^{n+\half}$. Using these values, the concentration at $t^{n+1}$ is then computed with a midpoint corrector, and the velocity $u^{n+1}$ is determined from midpoint and trapezoidal source terms.
%We predict concentrations at $t^{n+\half}$ and then use a midpoint corrector scheme to compute the final %concentrations at $t^{n+1}$.
%We predict velocity at $t^{n+1,*}$ and later correct this field using midpoint and trapezoidal source %terms to compute the final velocity at $t^{n+1}$.
More details can be found in \citenum{donev2019fluctuating}, but the main steps are summarized here; note the discretizations for the spatial gradients are not included for ease of presentation. \\ \\
%\Klymko{Note for self: use Fickian diffusion form below}}
%At the beginning of the time step, $t^n$, we have $c^n$ and $u^n$ and follow these steps:\\ \\
{\bf Step 1:}
Compute the predictor species fluxes as
\begin{eqnarray}
\SpeciesFlux^n &=& \rho_0 D \left [ \nabla c +  c (1-c)
\frac{\partial^2 g^{\rm ex}(c)}{\partial c^2} \nabla c +  c(1-c) \kappa \nabla \nabla^2 c \right]^n +
\left[\frac{2 z m c (1-c)}{k_B T} \right]^n \nabla \phi^n  \nonumber \\
&&+
\left( \sqrt{\frac{2\rho_0 m}{\Delta t \Delta V} D c(1-c)}\right)^n \WhiteNoiseMass^n 
%\left( \sqrt{\frac{2k_{\text{B}}}{\Delta t \Delta V} 2 D c (1-c)} \right)^n \left(\WhiteNoiseMomentum^{\text{mass}}_{(1)}\right)^n \ ,
\end{eqnarray}
where %$(\WhiteNoiseMomentum^{\text{mass}}_{(1)})^n$ \Carney{where 
$\WhiteNoiseMass^n$ are the i.i.d.~normal random variables and the electric potential is computed by solving the Poisson equation
\begin{equation}
    -\nabla \cdot \left( \epsilon \nabla \phi^n \right) = \left(q^f\right)^n
\end{equation}
with a cell-centered multigrid solver.
Compute the predictor reversible stress tensor as
\begin{equation}
    \ReversibleStress^n = \frac{\rho_0 k_B T}{m} \left[ \frac{\kappa}{2} |\nabla c|^2 \mathbb{I} - \kappa \nabla c \otimes \nabla c \right]^n \ .
\end{equation}
%\Klymko{do we actually define $g_o$ in the paper? or only in the reversible stress definition?}
{\bf Step 2:}
Compute the predictor velocity and pressure, $u^{*,n+1}$ and $\pi^{*,n+\frac{1}{2}}$, by solving the linear, saddle-point Stokes system \cite{cai:2014}:
\begin{eqnarray}
    \frac{\rho_0 u^{*,n+1}-\rho_0 u^n}{\Delta t} + \nabla \pi^{*,n+\frac{1}{2}} &=& -\nabla \cdot (\rho_0 u u^T)^n +\frac{1}{2} \left( \eta \nabla^2 u^n + \eta \nabla^2 u^{n+1,*} \right)  \\
&+& \nabla \cdot  \sqrt{\frac{\eta k_{\text{B}}T}{\Delta t \Delta V}} \left( \WhiteNoiseMomentum+ \WhiteNoiseMomentum^T \right)^n + \left[ \nabla \cdot \left( \epsilon \nabla \phi \right) \nabla \phi\right]^n + \nabla \cdot \ReversibleStress^n \nonumber \\
\nabla \cdot u^{*,n+1} &=& 0,
\end{eqnarray}
where $\Delta V$ is the volume of a grid cell.\\ \\
{\bf Step 3:} Compute the predictor concentration $c^{*,n+\half}$ from
\begin{equation}
    \rho_0 c^{*,n+\half} = \rho_0 c^n+\frac{\Delta t}{2} \left( - \nabla \cdot \SpeciesFlux^n \right) - \frac{\Delta t}{2} \nabla \cdot \rho_0 c^n \left(\frac{ u^n + u^{*,n+1}}{2}\right)  \ .
\end{equation}
{\bf Step 4:}
Compute the corrector species fluxes as
\begin{eqnarray}
\SpeciesFlux^{*,n+\half} &=& \rho_0 D \left [ \nabla c +  c (1-c)
\frac{\partial^2 g^{\rm ex}(c)}{\partial c^2} \nabla c +  c(1-c) \kappa \nabla \nabla^2 c \right]^{*,n+\half} \\
&&+ \left[\frac{2 z m c (1-c)}{k_B T} \right]^{*,n+\half} \nabla \phi^{*,n+\half} +
\left( \sqrt{\frac{2\rho_0 m}{\Delta t \Delta V} D c(1-c)}\right)^{*,n+\half}
\left( \frac{\left(\WhiteNoiseMass_{(1)}\right)^n+\left(\WhiteNoiseMass_{(2)}\right)^n}{\sqrt{2}} \right) ,\nonumber
\end{eqnarray}
where $\nabla \phi^{*,n+\half}$ comes from the multigrid solution to
\begin{equation}
    -\nabla \cdot \left( \epsilon \nabla \phi^{*,n+\half} \right) = \left(q^f\right)^{*,n+\half} ,
\end{equation}
and compute the corrector reversible stress tensor as
\begin{equation}
    \ReversibleStress^{*,n+\half} = \frac{\rho_0 k_B T \kappa}{m}\left[ \frac{1}{2} |\nabla c|^2 \mathbb{I} - \nabla c \otimes \nabla c \right]^{*,n+\half} \ .
\end{equation}
{\bf Step 5:}
Compute the corrector concentration $c^{n+1}$
\begin{equation}
    \rho_0 c^{n+1} = \rho_0 c^n+\Delta t \left( - \nabla \cdot \SpeciesFlux^{*,n+\half} \right) - \Delta t \, \nabla \cdot \rho_0 c^{*,n+\half} \left(\frac{ u^n + u^{*,n+1}}{2} \right) \ ,
\end{equation}
{\bf Step 6:}
Finally, compute the corrector velocity and pressure, $u^{n+1}$ and $\pi^{n+\half}$, by solving the Stokes system
\begin{eqnarray}
    \frac{\rho_0 u^{n+1}-\rho_0 u^n}{\Delta t} + \nabla \pi^{n+\frac{1}{2}} &=& -\frac{1}{2}\nabla \cdot \left [(\rho_0 u u^T)^n + (\rho_0 u u^T)^{*,n+1} \right] +\frac{1}{2} \left( \eta \nabla^2 u^n + \eta \nabla^2 u^{n+1} \right)  \\
&+& \nabla \cdot  \sqrt{\frac{\eta k_{\text{B}}T}{\Delta t \Delta V}} \left( \WhiteNoiseMomentum+ \WhiteNoiseMomentum^T \right)^n + \left[ \nabla \cdot \left( \epsilon \nabla \phi \right) \nabla \phi \right]^{*,n+\half} + \nabla \cdot \ReversibleStress^{*,n+\half} \nonumber \\
\nabla \cdot u^{n+1} &=& 0.
\end{eqnarray}

\section{Simulation Results}\label{sec:simulation_results}

\subsection{Parameter Calibration}\label{sec:calibrate}

To calibrate the model parameters, we select a specific RTIL that has
been studied extensively both experimentally and with molecular dynamics, namely, 
1-butyl-3-methylimidazolium hexafluorophosphate
or [DMPI+][F6P-].
Properties of [DMPI+][F6P-] (also known as [$\mathrm{C}_4\mathrm{min}^+$][$\mathrm{PF}_6^-$]) are summarized in Table \ref{tab:rtils}.
From the data in this table we can define the parameters needed by the code as summarized
in Table \ref{tab:properties}.
%\MarginPar{is relative permittivity a better term than dielectric constant -- JBB}

\begin{table}[h]
    \centering
            \caption{RTIL properties of [DMPI+][F6P-] at standard conditions from NIST database \texttt{https://ilthermo.boulder.nist.gov/}; ion diffusion coefficients from \citenum{IonDiffusionData1,IonDiffusionData2}.}\begin{tabular}{|c|c|c|c|}
        \hline\hline
          PubChem CID & CAS ID & Mass (g/mol) & Density (g/$\mathrm{cm^{3}}$)\\
         \hline
         2734174 & 174501-64-5 & 284.19 & 1.38 \\
        \hline
        Viscosity (cP) & Conductivity ($\mathrm{C~s^2~g^{-1}~cm^{-3}}$) & Relative Permittivity (-) & Sound speed (cm/s) \\
        \hline
        272    & $1.5\times 10^{-10}$  & 10.2 $\pm$ 0.4 & 144000 \\
        \hline
        Cation $D$ ($\mathrm{cm^2/s}$) & Anion $D$ ($\mathrm{cm^2/s}$) & Melting $T$ (K) & Entropy (J/(mol~K)) \\
        \hline
        $5.5 \times 10^{-8}$ & $4.3 \times 10^{-8}$ & 282 & 493 \\
        \hline\hline
    \end{tabular}
    \label{tab:rtils}
\end{table}

\begin{table}[h]
    \centering
   \caption{Simulation parameters.}
    \begin{tabular}{|c|c|}
    \hline\hline
    Density (g/cm$^3$) & $1.38 $ \\
    \hline
    Molecular mass (anion and cation) (g) & $2.4 \times 10^{-22}$ \\
    \hline
    Temperature (K)& 300\\
    \hline
    Charge per mass (C/g) & $6.8\times 10^2 $ \\
    \hline
    Relative Permittivity (-) & $10.4$ \\
    \hline
    Binary diffusion coefficient (cm/s$^2$) & $5.\times 10^{-8}$\\
    \hline
    Viscosity (cP) & $270.$ \\
    \hline\hline
        \end{tabular}
    \label{tab:properties}
\end{table}

To complete the specification of the model it remains to specify the 
excess Gibbs free energy, $g^{\rm ex}(c)$ and the interfacial tension parameter, $\kappa$.
Experimental measurements and molecular dynamics simulations show that the repulsive forces between cations and anions are strong enough to overcome the electrostatic forces and induce phase separation, where the morphological details depend on the specific
ionic liquid under consideration.  In the model, this repulsive force is represented by the excess free energy.  From a mathematical perspective, phase separation corresponds to an instability of the system. To assess this instability, we consider the linearized form of the concentration equation.
%In the absence of external fields, one expects that the cations
%and anions of a given RTIL will separate into distinct phases with features
%of a length scale specific to the given liquid 

For the case considered here, where the cations and anions are of equal mass, the concentration equation 
linearized around $c=\half$ must be unstable for the phases to separate.
The equation for a perturbation $\delta c$ about
one half is
\begin{equation}
\delta c_t = D \left [ \left( 1+ \frac{1}{4} \frac{\partial^2 g^{\rm ex}(c)}{\partial c^2}\mid_{c=\half} \right ) \nabla^2 \delta c - 
\frac{1}{4} \kappa \nabla^4 \delta c +
\frac{ z m }{2 k_B T} \nabla^2 \phi  \right )
\label{eq:delta_c_evolution}
\end{equation}
Observing that
\begin{equation}
-\epsilon \nabla^2 \phi = 2 \rho_0 z \delta c
\end{equation}
we then obtain
\begin{equation}\label{eq:delta_c_evolution_modified}
\delta c_t = D \left [ \left(
1+ \frac{1}{4} \frac{\partial^2 g^{\rm ex}(c)}{\partial c^2}\mid_{c=\half} \right ) \nabla^2 \delta c - 
\frac{1}{4} \kappa \nabla^4 \delta c -
\frac{\rho_0 z^2 m }{\epsilon k_B T} \delta c  \right ].
\end{equation}
%Note \MarginPar{Garcia, please look at this}
%\Carney{is this worth noting here, or is it extraneous to the current derivation?}\Garcia{Can move it to later. It's useful since $n = 1/\mathrm{(Ion~volume)}$} that for a 1:1 ionic liquid (i.e., cation and anion of valence 1),
%\begin{equation}
%\frac{\rho  z^2 m}{ \epsilon k_B T} = \frac{1}{\lambda_D^2} = 4 \pi \lambda_B n
%\end{equation}
%where $\lambda_D$ is the Debye length, $\lambda_B$ is the Bjerrum length and $n=\rho/m$ is the number density. For water the dielectric constant is $\epsilon_r \approx 80$ and $\lambda_B \approx 0.7$~nm while in ionic liquids $\epsilon_r \approx 10$ so $\lambda_B \approx 5$~nm \Carney{is this for \textit{all} ionic liquids? all RTILs? or just the RTILs we simulated?}\Garcia{Both RTILs and molten salts.}.
Taking the Fourier transform of \eqref{eq:delta_c_evolution_modified} gives
\begin{equation}
\deltachat_t = D \left [-\left(  \frac{1}{4} \frac{\partial^2 g^{\rm ex}(c)}{\partial c^2}\mid_{c=\half} +1 \right ) k^2  - 
\frac{1}{4} \kappa k^4 -
\frac{\rho_0 z^2 m }{ \epsilon k_B T}  \right ] \deltachat.
\label{eq:stability}
\end{equation}
where $k$ is the magnitude of the wave vector.
From this equation, one sees that both the electric field and the fourth-order term inhibit the growth of
perturbations and hence act to inhibit phase separation.
%In this equation, we see the stabilizing effect of the electric field 
For the system to be unstable, the 
coefficient of $\deltachat$ on the right hand side must be positive.
In general, this requires that the second derivative of $g^{\rm ex}$ be sufficiently negative and
%\MarginPar{mention $k^2$ overcoming stabilizing effect of electric field as well?}
$\kappa$ be sufficiently small for there to be a range of unstable $k$.
The $g^{\rm ex}$ term will then set the larger scale of the features, while
the $k^4$ term will regularize finer scale features.

The excess Gibbs free energy can be expressed in polynomial form \cite{Zhao_2020}; here we use,
\begin{equation}
g^{\rm ex}(c) = \alpha \left [ c (1-c) \right ] ^n
%\qquad\mathrm{and}\qquad
%g^{\rm ex}_i(x_i) = \frac12 \alpha \left [ x_i (1-x_i) \right ] %^n
\label{eq:gex}
\end{equation}
Experimental data indicates that the characteristic feature size 
of [DMPI+][F6P-]
is approximately 2-3 nm \cite{HardacreETAL_2010,HayeETAL_2015}.  
Accordingly, $g^{\rm ex}$ and $\kappa$ are chosen so that wavelengths
$\lambda = 2 \pi / k$ in the 4-6 nm range (twice the feature size)
are in the unstable range. 
%For $n=1$ the excess free energy has the form given by Flory-Huggins theory with $\alpha$ being the Flory $\chi$ parameter, which can be obtained from experiments \cite{yoo2012}.
From the parameters describing [DMPI+][F6P-], we can estimate the electric force
term in (\ref{eq:stability}) to be approximately $ 4 \times 10^{16}~\mathrm{cm}^{-2}$.
We choose $\alpha = 4.0 \times 10^3 \; \mathrm{cm}^{-1}\;\mathrm{s}^{-1}$ and $n=2$ in (\ref{eq:gex}) so the coefficient of $k^2$
in (\ref{eq:stability}) is approximately $1124 \;\mathrm{cm}^{-1}\;\mathrm{s}^{-1}$,
so that, ignoring interfacial tension, wavelengths shorter than approximately 10~nm are unstable.
Finally, for $\kappa = 1\times 10^{-11}~\mathrm{cm}^{-2}$ wavelengths between 3.3 and 9.4~nm are stable so we take this value as our baseline.  
% \MarginPar{But from the structure factor we say that the dominant length scale is 2~nm; does that length scale correspond to a half wavelength?? yes. a wavelength here is a full period of stripes. . .peak to peak to a single structure is half JBB }
It should be noted that we experimented with different forms for $g^{\rm ex}$ (different values of $\alpha$ and $n$) while maintaining the value of the
second derivative at $c=\half$ and found that the
specific form did not change the qualitative structure significantly.

\subsection{Bulk Morphology}\label{sec:BulkMorf}
\begin{figure}[h]
    \centering
    \includegraphics[width=.2\textwidth]{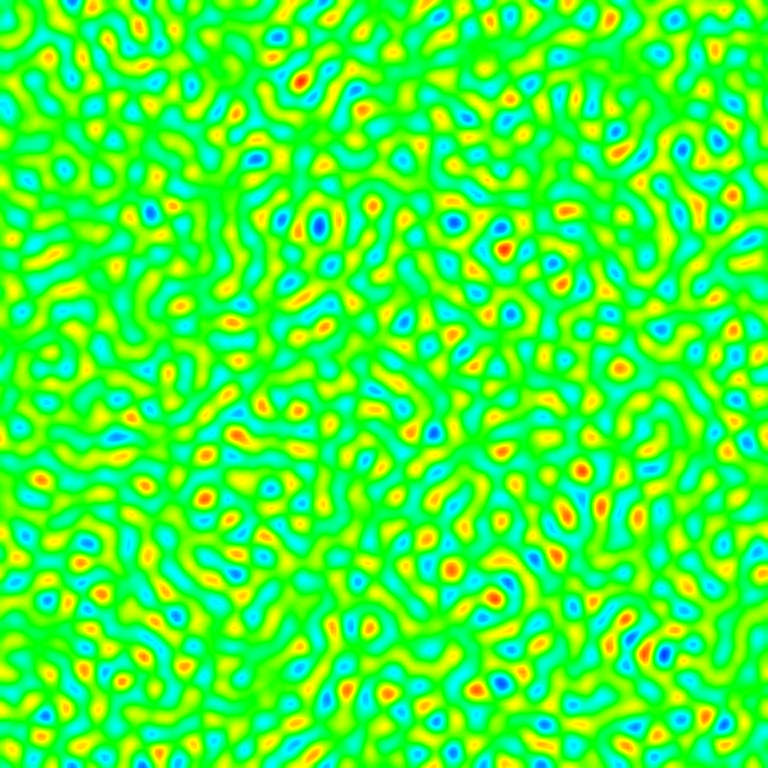}
    \includegraphics[width=.2\textwidth]{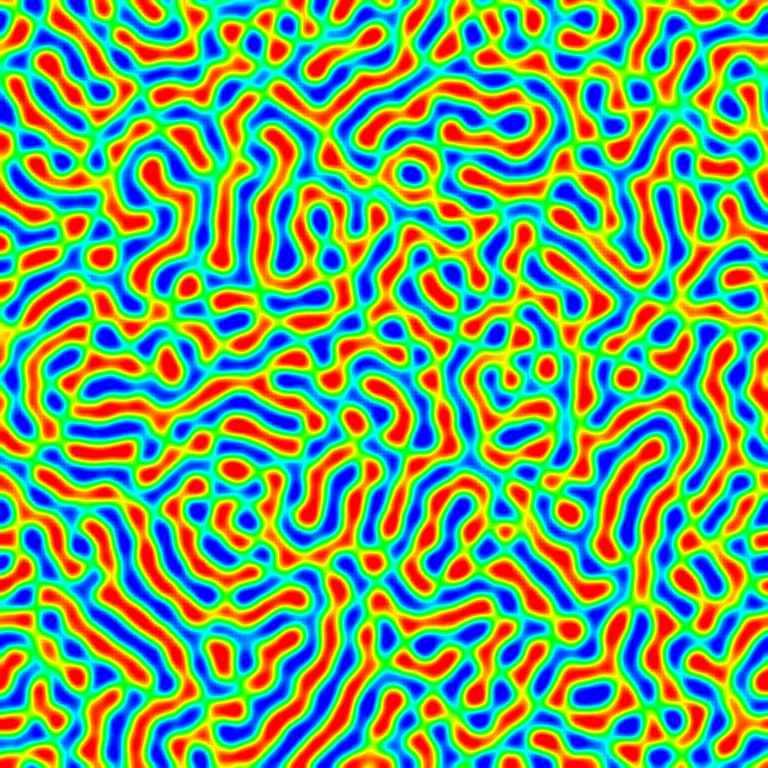}
    \includegraphics[width=.2\textwidth]{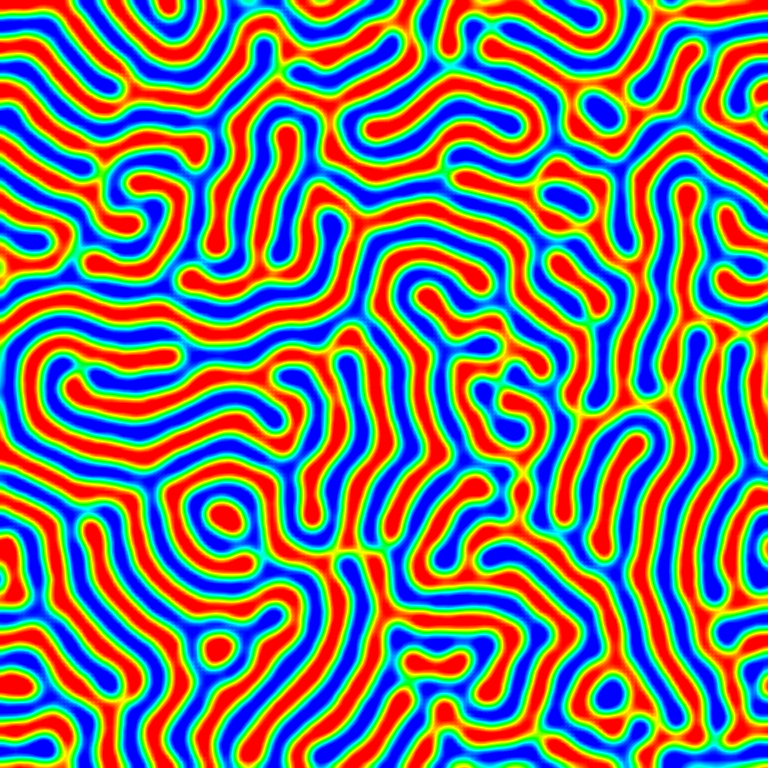}
    \includegraphics[width=.2\textwidth]{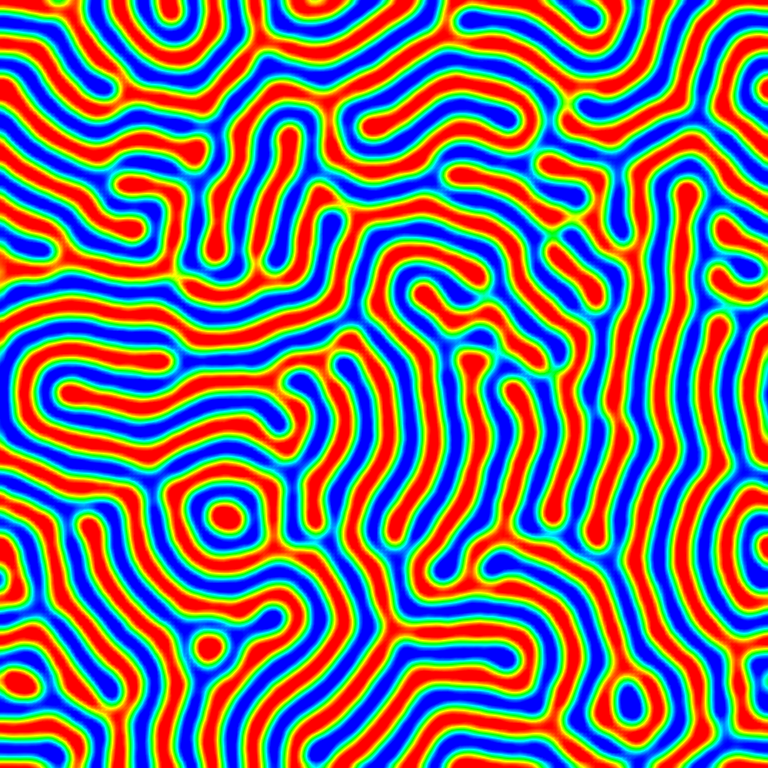}
    \includegraphics[width=.05\textwidth]{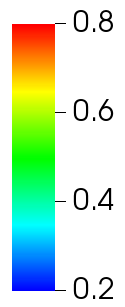}
    \caption{Time evolution of cation concentration profiles illustrating structural pattern formation at $t=1, 2, 5,$ and $10$~ns for $\kappa = 1\times 10^{-11}~\mathrm{cm}^{-2}$.  All figures in this paper use the same colorbar unless otherwise noted.}
    \label{fig:kc1d-11}
\end{figure}

First we consider quasi-two-dimensional systems with periodic boundary conditions.
All quasi-two-dimensional simulations in this paper use $192 \times 192 \times 1$ cells with grid spacings $\Delta x = \Delta y = \Delta z = 0.5$~nm and time step $\Delta t = 0.2~\mathrm{ps}$.
As discussed in Section \ref{sec:calibrate}, we take $\kappa = 1\times 10^{-11}~\mathrm{cm}^{-2}$ for our baseline case; see Table~\ref{tab:properties} for other parameters. Figure~\ref{fig:kc1d-11} shows the development of patterns that form in a simulation of the RTIL starting from a homogeneous initial condition of $c=\half$.  By $t=5$~ns, the morphology nearly reaches the final configuration we show at $t=10$~ns.  
In fact, at a later time of $t=50$~ns (not pictured), the morphology is nearly identical to the $t=10$~ns frame.  
This stable feature size reflects the competition between the phase separation and electrostatic forces as discussed in Section \ref{sec:calibrate}.
Also, note that all figures in this paper use the same colobar for cation concentration used in Figure~\ref{fig:kc1d-11} unless otherwise noted.

\begin{figure}[h]
    \centering
    \includegraphics[width=.2\textwidth]{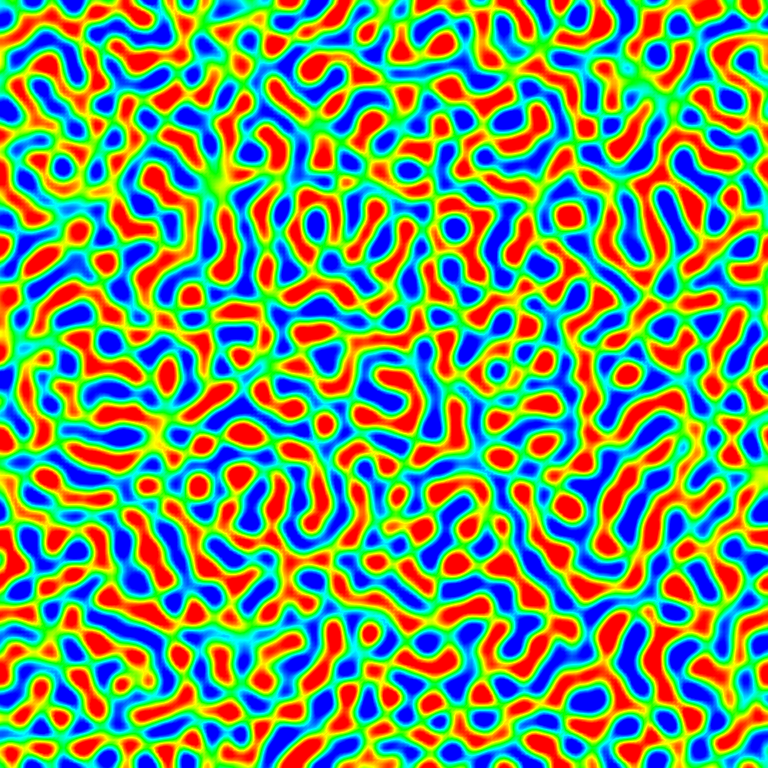}
    \includegraphics[width=.2\textwidth]{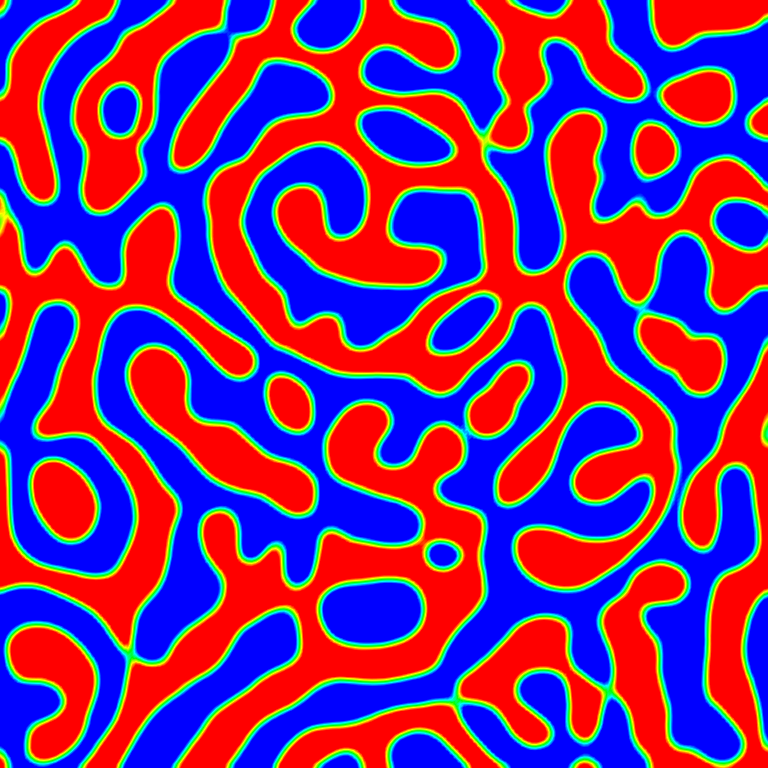}
    \includegraphics[width=.2\textwidth]{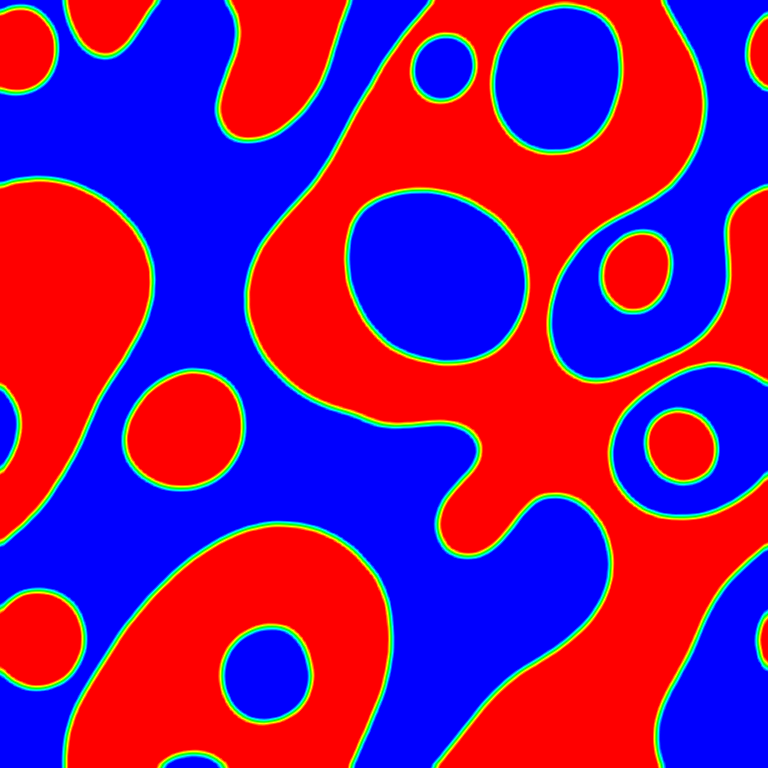}
    \includegraphics[width=.2\textwidth]{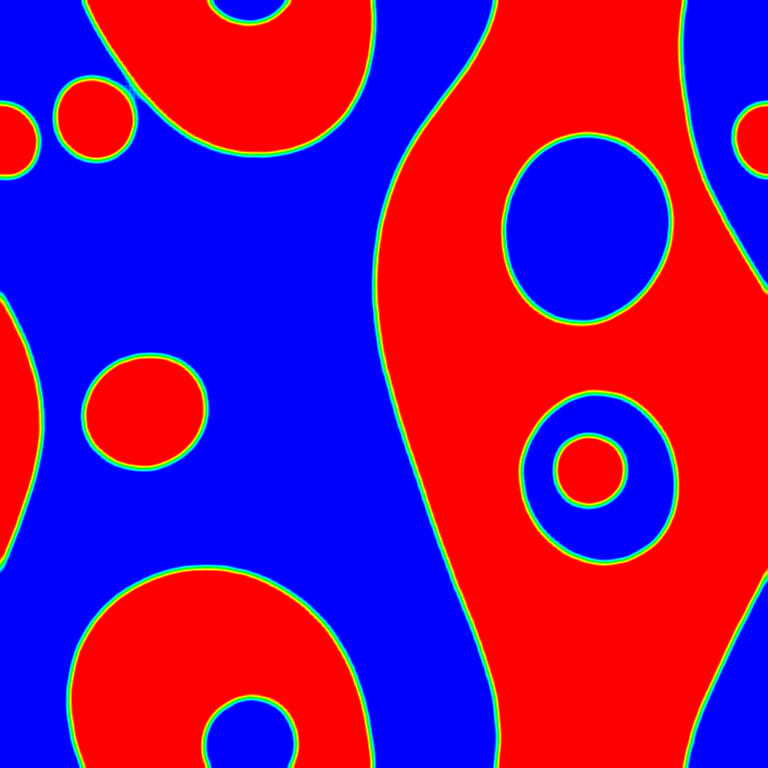}
    \caption{Time evolution of pattern formation as in Figure \ref{fig:kc1d-11} with uncharged species ($z = 0$) at $t=1, 2, 5,$ and $10$~ns.  
    In this case the regions will continue to coarsen.}
    \label{fig:uncharged}
\end{figure}

\begin{figure}[h]
    \centering
    \includegraphics[width=.3\textwidth]{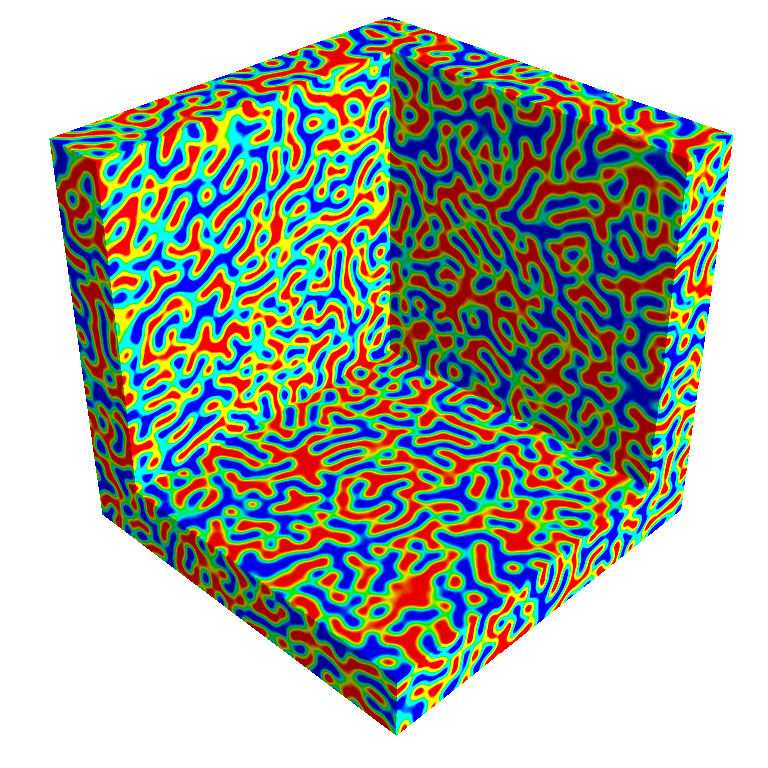}
    \includegraphics[width=.3\textwidth]{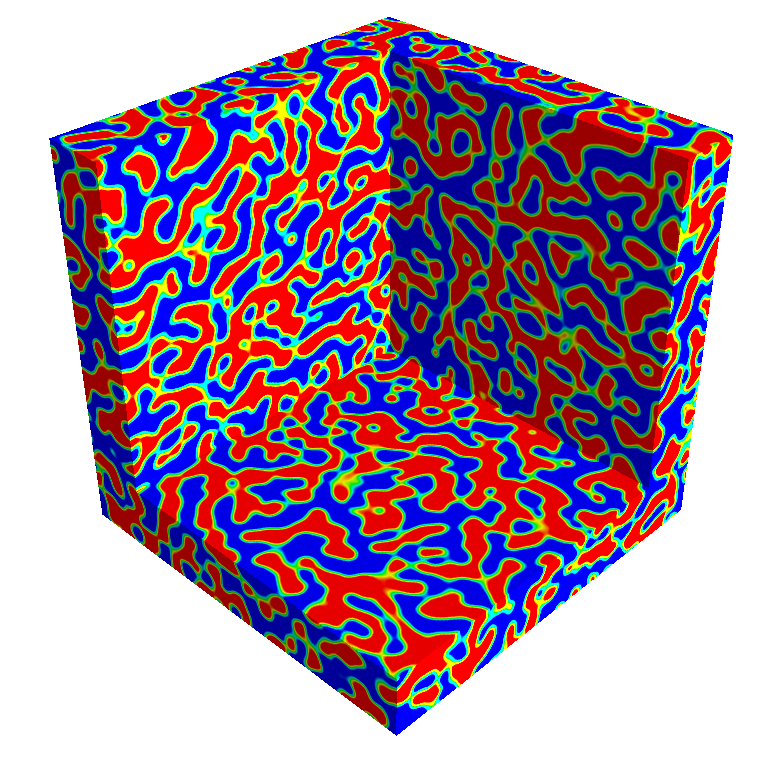}
    \includegraphics[width=.3\textwidth]{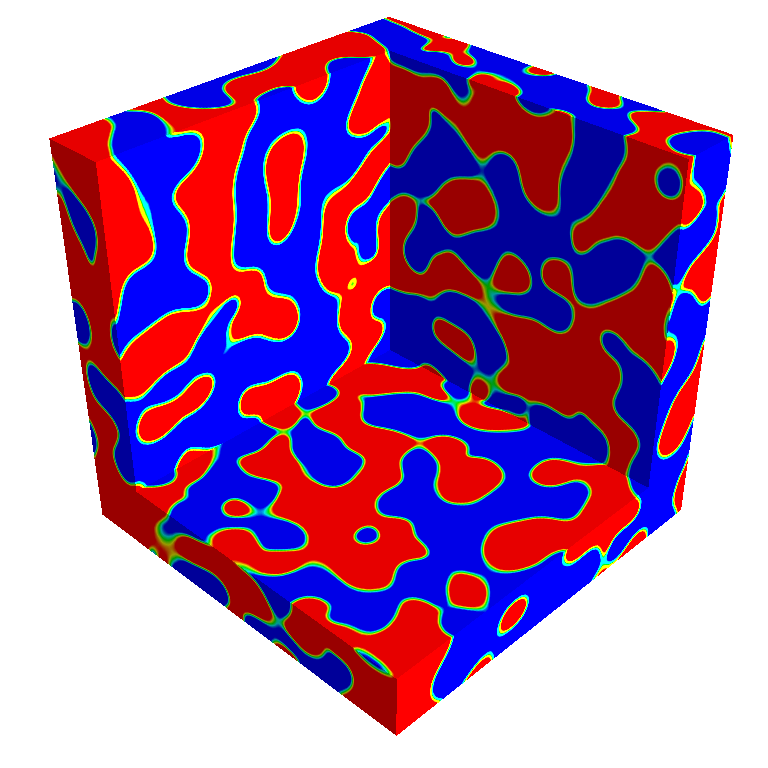}
    \caption{Cation concentration profiles from three-dimensional simulations (interior cut-out image). The left panel shows the stable pattern ($t=10$~ns) reached by the RTIL; compare with the last frame in Figure \ref{fig:kc1d-11}. The other two panels show the evolution of pattern formation of the cation concentration in three dimensions for the uncharged case at $t=2$~ns and $t=5$~ns; compare with the second and third frames in Figure \ref{fig:uncharged}.}
    \label{fig:3D}
\end{figure}

In Figure \ref{fig:uncharged} we show the analogous spinodal decomposition as in Figure \ref{fig:kc1d-11} but for the case of uncharged species ($z=0$); here the patterns coarsen quickly and increasingly with time.
%\MarginPar{Whenever possible the times shown in Figs. 1, 2, 3, and 4 should match. -- Garcia.   JBB:  end of 1 and 4 are at same time.  i don't think we need to run spinodal things with charges out that long.}
We repeat the simulations, but in three dimensions (see Figure~\ref{fig:3D}) using a cubical domain of $192 \times 192 \times 192$ cells with the same mesh spacing and time step as before.
As in two dimensions, the case with charges evolves to a pattern with fixed feature sizes and then stabilizes whereas the uncharged case coarsens quickly and continues to coarsen over time.

\begin{figure}[h]
    \centering
    \includegraphics[width=.2\textwidth]{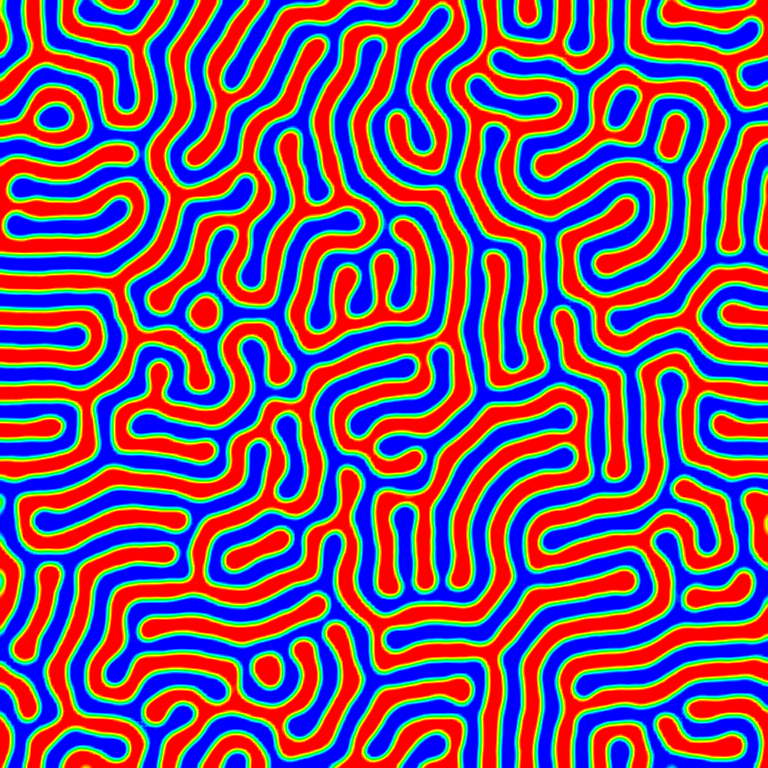}
    \includegraphics[width=.2\textwidth]{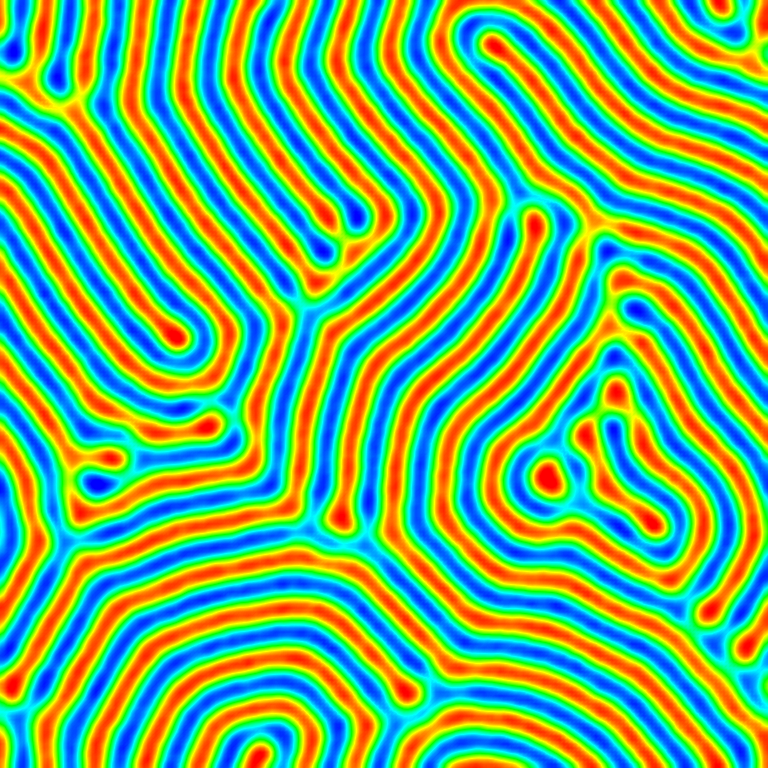}
    \includegraphics[width=.2\textwidth]{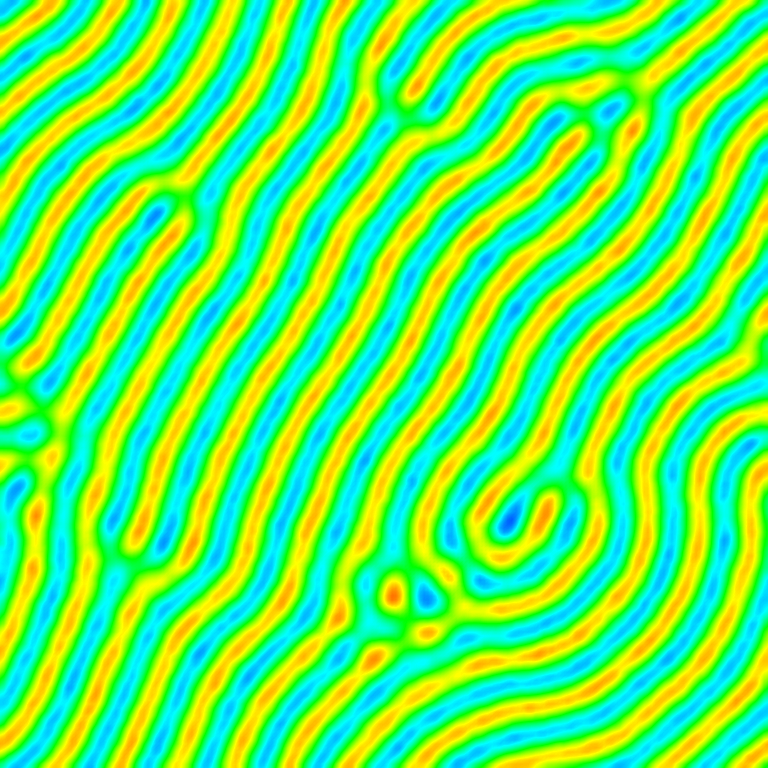}
    \includegraphics[width=.2\textwidth]{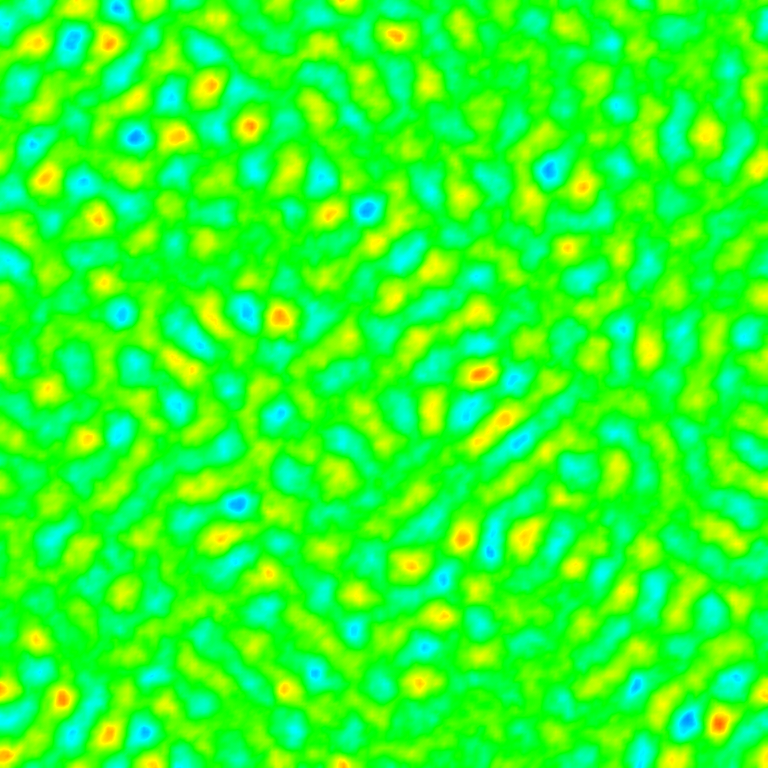}
    \caption{Cation concentration profiles illustrating spinodal decomposition for the $t=50$~ns configuration for $\kappa=0.5, 1.5, 2.0,$ and $2.5\times 10^{-11}~\mathrm{cm}^{-2}$, respectively.  Compare to final image in Figure \ref{fig:kc1d-11}.  We note that the colorbar for the $\kappa=2.5\times 10^{-11}~\mathrm{cm}^{-2}$ case only has been changed to range from 0.4 to 0.6 in order to see the structure.}
    \label{fig:kc_range}
\end{figure}

Next we examine how the morphology depends on the interfacial tension parameter, $\kappa$.  Figure \ref{fig:kc_range} shows the patterns formed at four additional values of $\kappa$. 
As $\kappa$ increases, the dynamic range of phase separation concentrations decreases.  For the largest value shown ($\kappa=2.5\times 10^{-11}~\mathrm{cm}^{-2}$) the phase separation is almost completely suppressed. 

\begin{figure}[h]
  \centering
  \includegraphics[width=.425\textwidth]{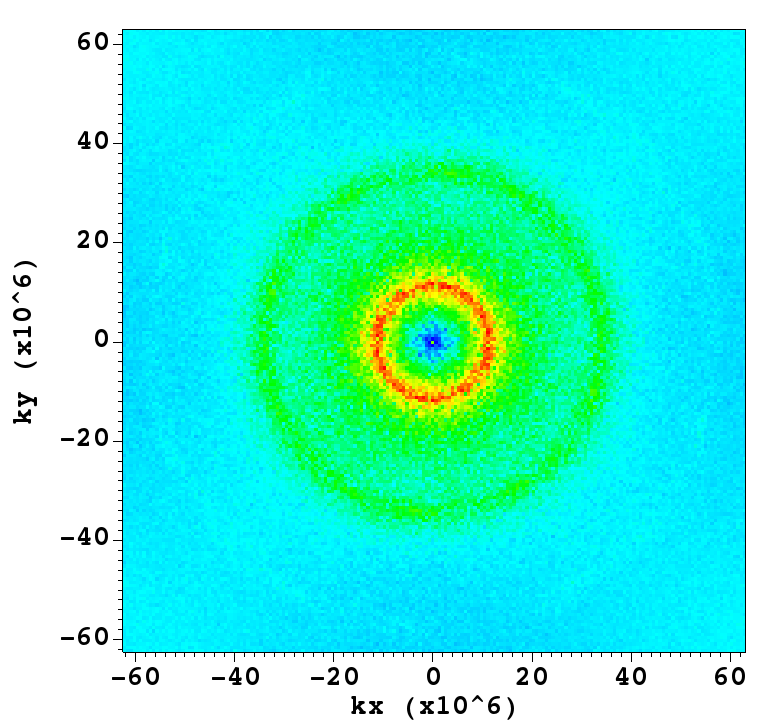}
  \includegraphics[width=.425\textwidth]{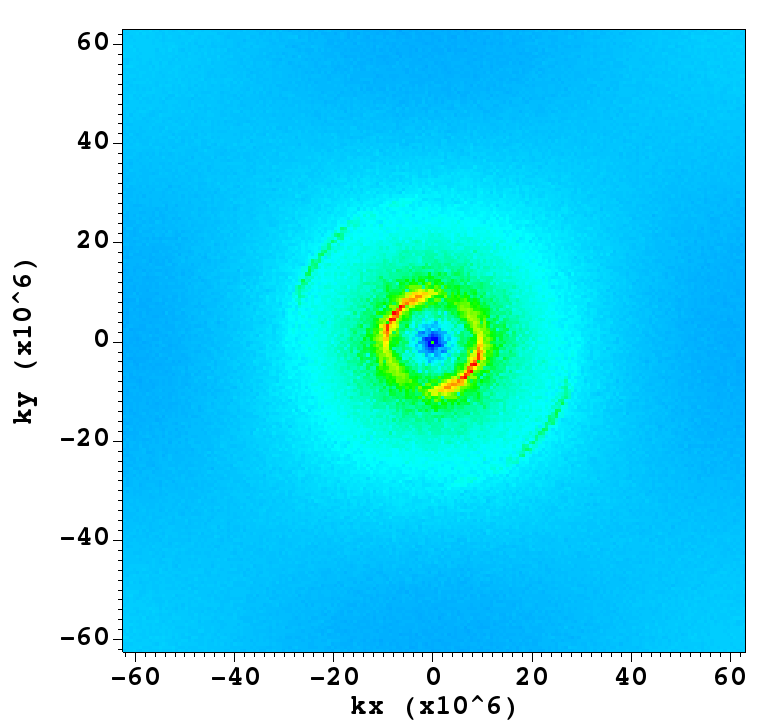}
  \includegraphics[width=.1\textwidth]{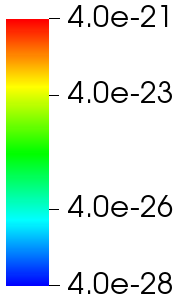}
  \caption{Structure factor in logscale for the cation concentration, $S(\boldsymbol{k})$ for %$$\kappa=0.5\times 10^{-11}~\mathrm{cm}^{-2},
  $\kappa = 1\times 10^{-11}~\mathrm{cm}^{-2}$ (left) and $\kappa = 2\times 10^{-11}~\mathrm{cm}^{-2}$ (right).
  The radius of the ring, $k_r$, is inversely proportional to the morphological feature size as $\ell \sim~\pi/k_r$.
  The anisotropic ring in the right panel is consistent with the striped features favoring a single orientation in physical space, as seen in the third panel in Figure \ref{fig:kc_range}.}
  \label{fig:SF}
\end{figure}

In addition, as $\kappa$ increases, the feature size in these patterns becomes larger.
To quantify this observation we measured the
static structure factor, which is the Fourier transform of the equal time covariance of the concentration,
\begin{equation}
S(\boldsymbol{k}) = \left\langle \left(\widehat{\delta c_{\boldsymbol{k}}}\right) \left(\widehat{\delta {c}_{\boldsymbol{k}}}\right)^* \right\rangle,
\end{equation}
where the brackets $\langle ~\rangle$ denote an equilibrium average over time and 
\begin{equation}
\widehat{\delta c_{\boldsymbol{k}}}(t) = \int \delta c (\boldsymbol{x},t)~e^{-i\boldsymbol{k}\cdot \boldsymbol{x}}d\boldsymbol{x}.
\end{equation}
In each case, we capture statistics for the structure factor by sampling at every time step for 1~ns beginning at $t=50$~ns.
Figure~\ref{fig:SF} shows the structure factor for the $\kappa = 1.\times 10^{-11}~\mathrm{cm}^{-2}$ and $\kappa = 2.\times 10^{-11}~\mathrm{cm}^{-2}$ cases.
For each value of $\kappa$, the structure factor has a maximum at a radius of $|\boldsymbol{k}| = k_r$.
We note that for $\kappa = 2.\times 10^{-11}~\mathrm{cm}^{-2}$, we see only a partial ring that reflects the fact that most of the stripes in Figure \ref{fig:kc_range} are oriented in the same direction.
The corresponding patterns feature a length scale (i.e., the width of the red or blue structures) that can be found using $\ell = \pi/k_r$.
We compute $k_r$ by considering $S(\boldsymbol{k})$ to be a probability density function and computing its expected value,
\begin{equation}
k_r = \frac{\sum_{k_x}\sum_{k_y}\sqrt{k_x^2+k_y^2}~S(k_x,k_y)}{\sum_{k_x}\sum_{k_y}S(k_x,k_y)},
\end{equation}
where we only include points in the sum where $S(\boldsymbol{k})$ is above 1\% of the peak value,
%\Garcia{Is it "within" or "above"?}
which effectively acts as a white noise filter.

Table \ref{table:SF} lists the values of $k_r$ and $\ell$ for all five values of $\kappa$ that we consider (the  cases not pictured in Fig.~\ref{fig:kc_range} have rings similar to the $\kappa = 1.\times 10^{-11}~\mathrm{cm}^{-2}$ case).
We note that the results for the baseline $\kappa = 1.\times 10^{-11}~\mathrm{cm}^{-2}$ case with $\ell=2.9$~nm are consistent with experimental measurements~\cite{Triolo2006,Triolo2007,HardacreETAL_2010} and molecular dynamics simulations~\cite{Morrow2002,Liu2004} for the RTIL discussed in Section \ref{sec:calibrate}; for a review see~\citenum{HayeETAL_2015}.
As $\kappa$ increases, the associated $\ell$ increases as well, which is consistent with the patterns in Figures \ref{fig:kc1d-11} and \ref{fig:kc_range}.
Finally, our three-dimensional simulations using $\kappa = 1.\times 10^{-11}~\mathrm{cm}^{-2}$ show a spherical structure factor (not pictured) with $k_r\sim 1.1\times 10^7$~cm, which matches the two-dimensional case.
%\MarginPar{technically, i think the wavelength for a given $k$ is $2 \pi / k$. feature are half of the wavelength.  AJN: yes I agree; modified text to reflect this}

\begin{table}[h]
\centering
\caption{The radius corresponding to the maximum value of the structure factor, $k_r$, and the associated pattern feature scale, $\ell=\pi/k_r$ as a function of $\kappa$.}
\begin{tabular}{|c|c|c|} 
 \hline
 \hline
 $\kappa$ $[10^{-11}$~cm$^{-2}]$ & $k_r$ [$\mathrm{nm}^{-1}$] & $\ell$ [nm] \\
 \hline
 0.5 & 1.35 & 2.33 \\
 1.0 & 1.14 & 2.76 \\
 1.5 & 1.03 & 3.05 \\
 2.0 & 0.96 & 3.27 \\
 2.5 & 0.92 & 3.41 \\
 \hline
 \hline
\end{tabular}
\label{table:SF}
\end{table}
%\clearpage

\subsection{Double Layer Capacitance}\label{sec:Capacitance}

Consider a charged fluid, either an electrolyte solution or an ionic liquid, confined in a parallel plate capacitor with electrodes at $y = 0$ and $y = L$.
The specific differential capacitance of the double layer is defined as $C = {d \sigma_s}/{d V}$ where $\sigma_s$ is the surface-charge density and $V$ is the overall potential drop between the bulk of electrolyte and the electrode surface. By Gauss' law, for the electrode wall at $y=0$,
\begin{align}
    \sigma_s = - \epsilon \left. \frac{d\phi}{dy}\right|_{wall}
\end{align}
With this,
\begin{align}
    C(V) = - \epsilon \frac{d}{d V} \left. \frac{d\phi}{dy} \right|_{wall}
    \label{eqn:Capacitance}
\end{align}
From Gouy-Chapman theory, for a 1:1 electrolyte solution, 
\begin{align}
    C_\mathrm{e} = \frac{\epsilon}{\lambda_D} \cosh{ \Phi } 
    \qquad \mathrm{where} \qquad \Phi = \frac{e V}{2 k_B T}
    \label{GouyChapmanCapacitance}
\end{align}
and
\begin{align}
    \lambda_D = \sqrt{\frac{\epsilon k_B T}{z^2 m \rho_0}}
\end{align}
is the Debye length. 

To account for steric effects in ionic liquids Kornyshev \cite{DoubleLayerCapacitance} formulated a lattice model with a lattice saturation parameter $\gamma$ defined as the ratio of the total number of ions to the number of available sites ($0 \leq \gamma \leq 1)$. 
This formulation was improved \cite{GoodwinETAL_2017} by accounting for the enthalpy of mixing contribution to the free energy, which adds another parameter, $\alpha$, to the model ($0 \leq \alpha \leq 1$). For this model the capacitance of the double layer for an ionic liquid with equal size ions is
\begin{align}
    C = \frac{\epsilon}{\lambda_D}~\frac{\sqrt{\alpha} \cosh(\alpha \Phi)}{1 + f} \sqrt{\frac{f}{\ln(1+f)}}
    \qquad \mathrm{where} \qquad
    f = 2 \gamma \sinh^2 (\alpha \Phi)
    \label{eqn:Kornyshev}
\end{align}
This reduces to the Gouy-Chapman result for $\alpha = 1$ in the limit $\gamma \rightarrow 0$.
Qualitatively the capacitance $C(V)$ has the so-called ``bell'' curve shape for large $\gamma$ while for small $\gamma$ the capacitance shows a dip near $V=0$ (``camel'' shape).
The former case is typical of ionic liquids for which the double layer thickness increases with voltage.
The latter case corresponds to dilute electrolyte solutions, where $dC/dV > 0$ for small voltages and the thickness decreases with voltage until steric effects become significant.

We measure the differential capacitance using a series of quasi-two-dimensional simulations.
Our simulations use a parallel plate capacitor geometry with no-slip, impermeable, fixed voltage walls in the $y$ direction, and periodic in the $x$ direction.
For each simulation, the voltage at the top and bottom walls are equal in magnitude but opposite in sign; otherwise the parameters used were identical to those used in the periodic simulations in Section \ref{sec:BulkMorf}.
We performed simulations using $\pm$1, 2, 4, 8, 16, 32, 64, and 128~V at the walls for the baseline case of $\kappa = 1.0 \times 10^{-11} \mathrm{cm}^{-2}$.

\begin{figure}[h]
    \centering
    \includegraphics[width=.2\textwidth]{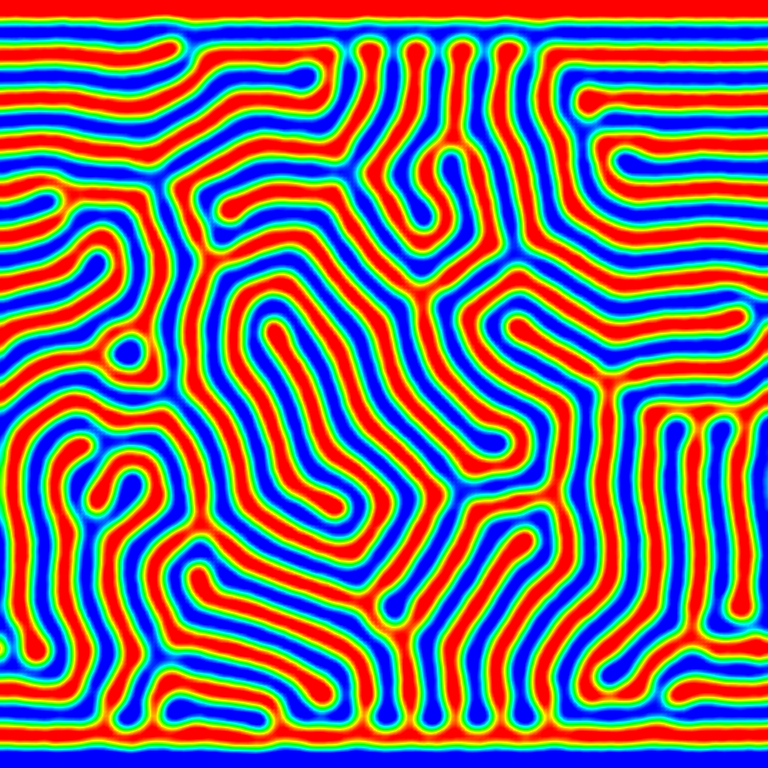}
    \includegraphics[width=.2\textwidth]{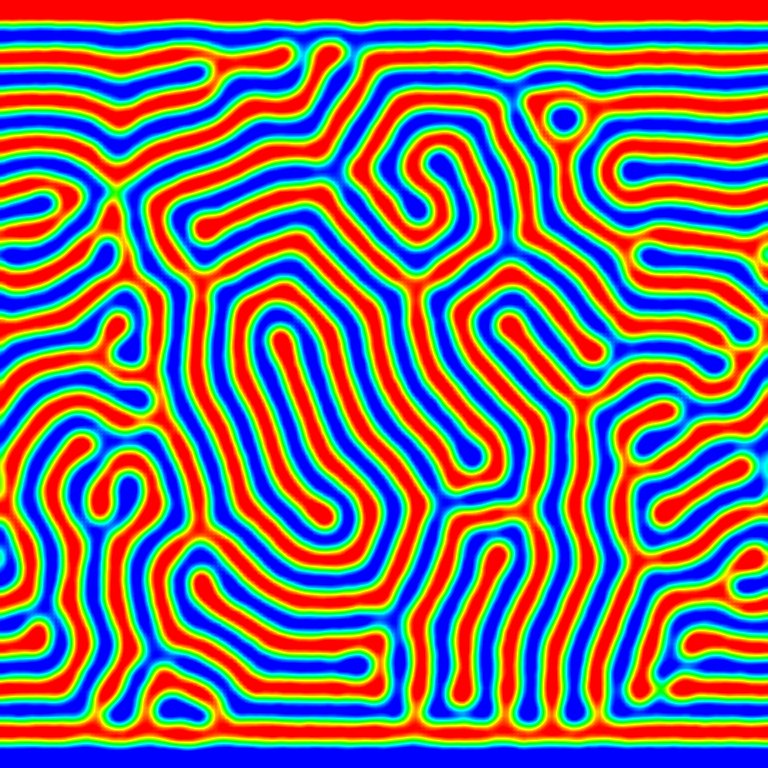}
    \includegraphics[width=.2\textwidth]{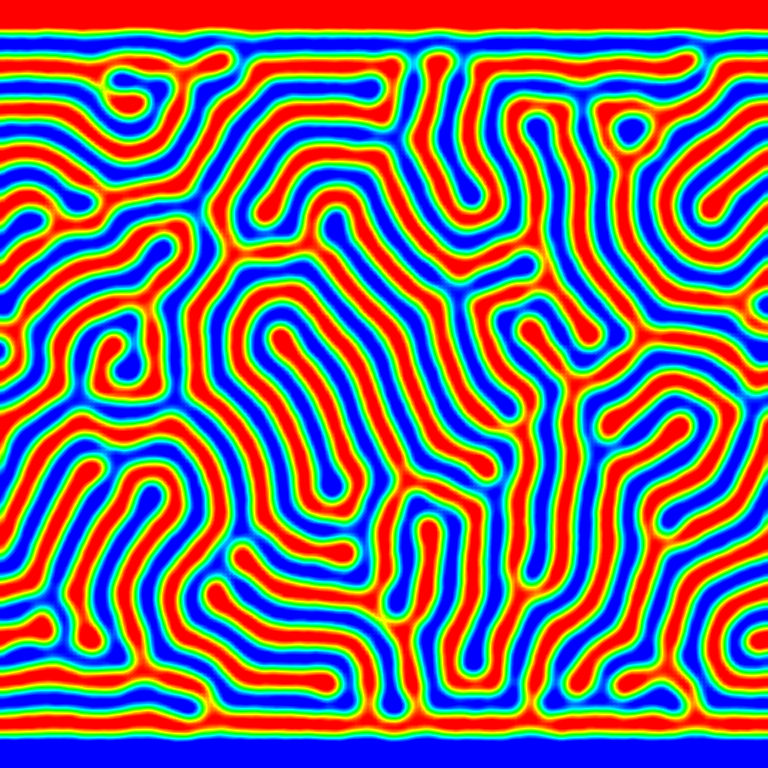}
    \includegraphics[width=.2\textwidth]{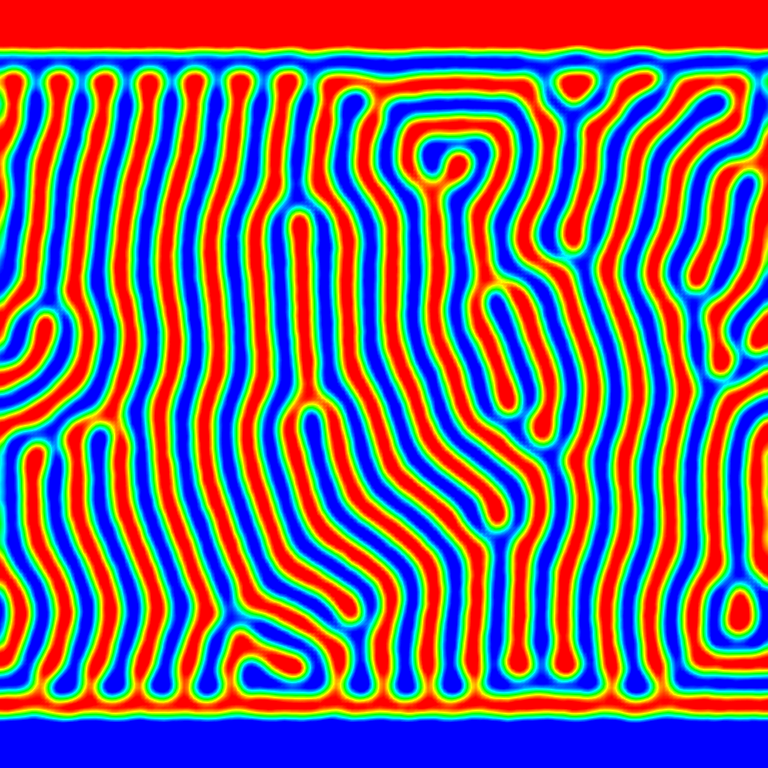}
    \caption{Cation concentrations for capacitance simulations. Electrode wall voltages (top and bottom) are $\pm$2, 8, 32, and 128V.}
    \label{fig:2D_capacitance}
\end{figure}

Figure \ref{fig:2D_capacitance} shows the cation concentration after equilibration for several of the cases with varying electrode voltages. The equilibration time depends on the voltage at the walls, with smaller electrode voltages taking longer to fully form the double layer structure, which is why for most of the simulations the voltage was above the electrochemical window of the RTIL \cite{Buzzeo2004}.
The simulations ran until equilibration, where the patterns had reached a steady configuration
(80~ns for runs with $|V|\ge 16$~V, up to 400~ns for the $|V|=1$ case).
%The double layer clearly increases with voltage. % Moved this statement below AG
Note that for the largest voltage presented (128V), the morphology features a vertical striped pattern parallel to the strong electric field.

\begin{figure}[h]
  \centering
  \includegraphics[width=.7\textwidth]{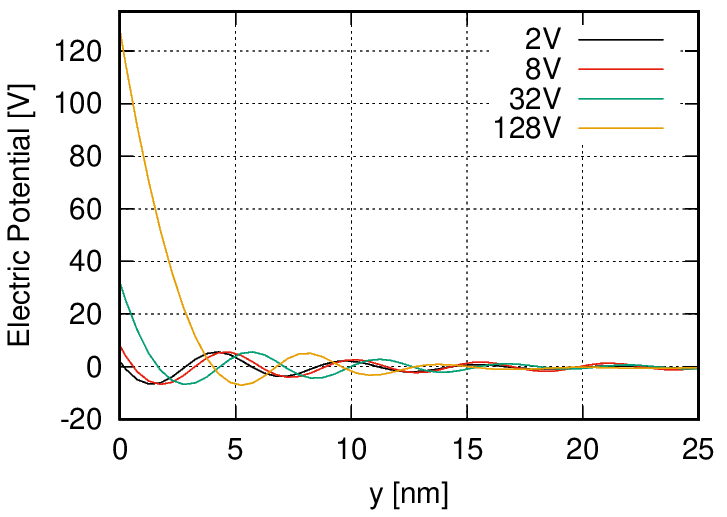}
  \caption{Horizontally-averaged potential for the capacitance cases in Figure \ref{fig:2D_capacitance}. }  
  \label{fig:normal}
\end{figure}

To calculate the capacitance from the simulation data we computed the surface charge density $-\epsilon d\phi/dy$ by horizontally averaging $\phi$ and then approximating the normal derivative at the wall using a second-order finite difference approximation using the boundary potential and two interior values.
In Figure \ref{fig:normal} we show plots of the horizontally-averaged electric potential for the cases depicted in Figure \ref{fig:2D_capacitance}.
First, we see that the normal derivative of $\phi$ steepens with increasing voltage.
The double layer thickness also clearly increases with voltage.
We also observe that the amplitude of the patterns a few nanometers away from the wall is similar for all voltages.

\begin{figure}
  \centering
  \includegraphics[width=.7\textwidth]{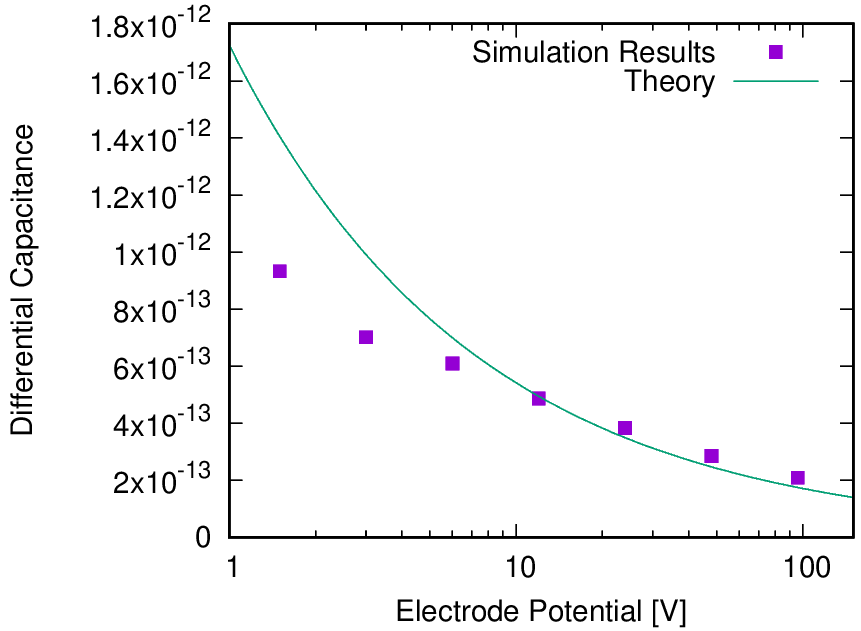}
  \caption{Differential capacitance as a function of electrode potential compared to theory (\ref{eqn:Kornyshev}).}  
  \label{fig:dcap}
\end{figure}

From these simulations we compute the differential capacitance using (\ref{eqn:Capacitance}) by estimating the derivative of the surface charge density with respect to voltage using a second-order finite difference approximation.
The measured differential capacitance as a function of electrode voltage is shown in Figure~\ref{fig:dcap}.
The simulation data was curve fit to (\ref{eqn:Kornyshev}) and the parameter values for the  optimal fit were $\gamma = 1.0$ and $\alpha = 1.0$; the corresponding curve is also shown in Figure \ref{fig:dcap}.
For larger voltages we are able to recover the predicted differential capacitance.
For smaller potentials our simulations under-predict the differential capacitance.

\begin{figure}[h]
    \centering
    \includegraphics[width=.3\textwidth]{final_images/cap_8V.png}
    \includegraphics[width=.3\textwidth]{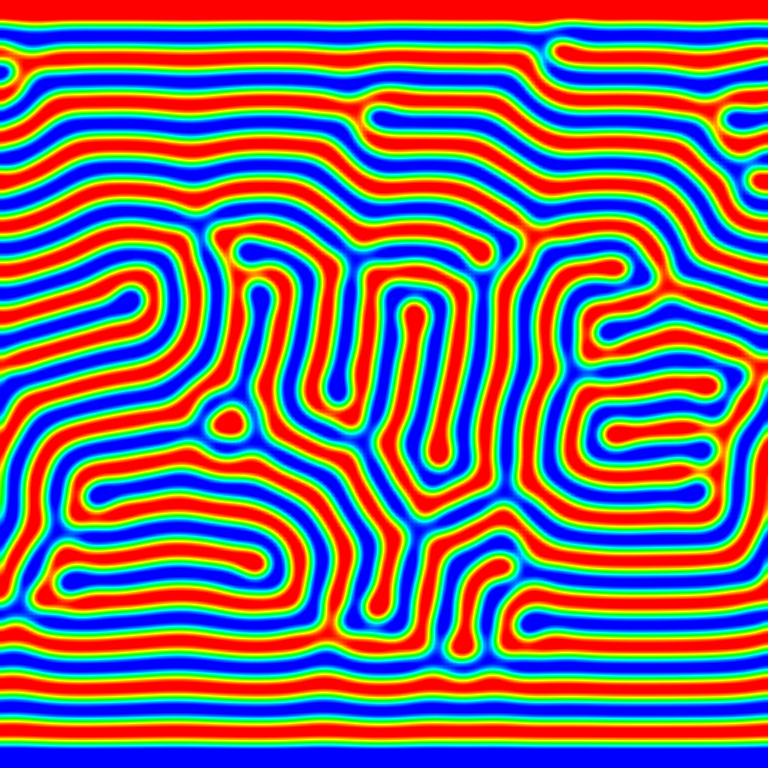}
    \includegraphics[width=.3\textwidth]{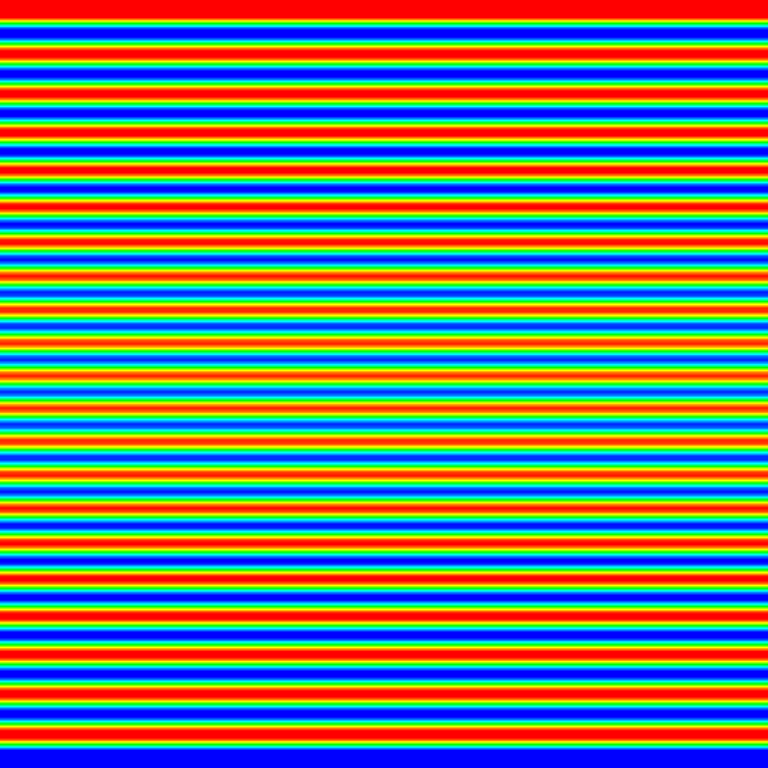}
    \caption{Cation concentrations for $\pm$8~V capacitance simulations. (Left) Stochastic simulation, (Middle) deterministic simulation with an initial stochastic perturbation, and (Right) purely deterministic simulation.}
    \label{fig:2D_capacitance_det}
\end{figure}

Next we illustrate the effect of the thermal fluctuations in the fluctuating hydrodynamic model of the RTIL. 
We consider two cases in which the simulations were performed with the stochastic fluxes turned off: a deterministic simulation with a random initial perturbation (first running one time step of the full stochastic algorithm and then turning off the noise terms); and a fully deterministic simulation with a homogeneous initial condition. For the periodic systems considered in Section \ref{sec:BulkMorf} we found little difference between the fully stochastic and the randomly perturbed deterministic simulations. 
% \Garcia{What does the pure deterministic simulation do? Does round-off break symmetry to form patterns or does it just "sit there"?. 
%AJN: purely deterministic periodic morphology sits absolutely still}
However, if we consider the steady state cation concentration in capacitance simulations with $\pm$8~V electrode potentials as shown in Fig.~\ref{fig:2D_capacitance_det}, we see that the results for the two deterministic cases are very different from the results for the fully stochastic simulation. 
The initially perturbed deterministic simulation has a similar morphology, except that horizontal stripes are preferred in the vicinity of the walls.
In the fully deterministic simulation, horizontal stripes quickly form across the domain. These stripes do not have a consistent structure size; they are thinner at the center of the domain and thickest near the walls.

\begin{figure}[h]
  \centering
  \includegraphics[width=.7\textwidth]{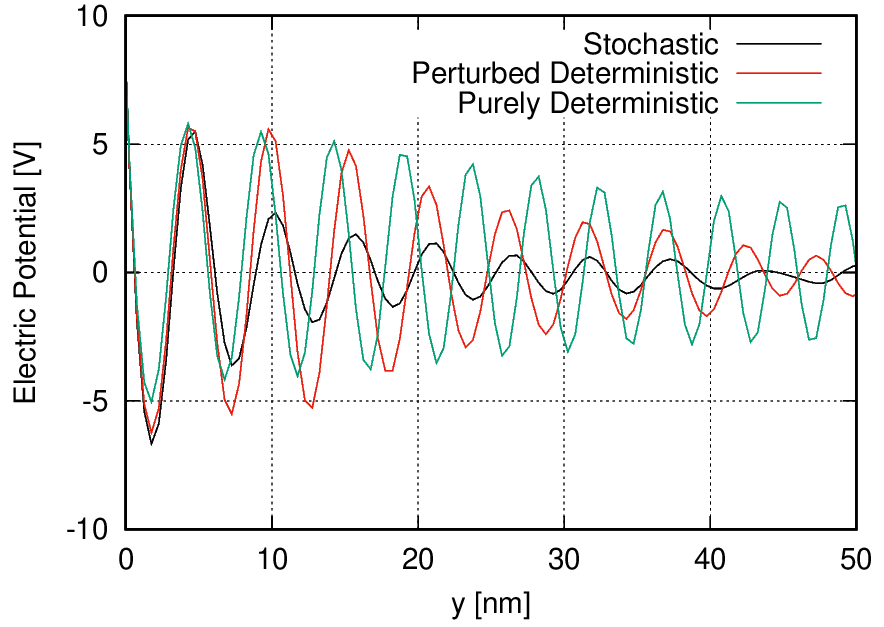}
  \caption{Horizontally-averaged potential for the cases in Figure \ref{fig:2D_capacitance_det}.  The perturbed deterministic case has a large amplitude near the walls due to the horizontal striping, but the same wavelength as the stochastic case.  The purely deterministic case has striping features with smaller wavelength.}
  \label{fig:normal_det}
\end{figure}

In Figure \ref{fig:normal_det} we show horizontally-averaged profiles of the electric potential for these same three cases which further confirm these observations.
The perturbed deterministic simulation shows a slower decrease in the potential away from the wall due to the horizontal striping (constructive interference in the horizontal averaging), and in the purely deterministic case the wavelength of the oscillations is clearly smaller.

\subsection{Electroosmotic shear flow}

Electroosmotic flow occurs when an external electric field is applied parallel to the electric double layer near an electrode wall. Since the double layer is not electroneutral the ions near the wall experience a Lorentz force, which results in a body force on the fluid.
For channel walls of equal potential (or surface charge density) electroosmosis results in a plug flow for wide channels ($L \gg \lambda_D$) and Poiseuille-type flow in narrow channels. Here we consider electrode walls of opposite potential, as in the parallel plate capacitor geometry described in the previous section, which results in electroosmotic shear flow. Specifically we consider electrodes
with potentials of $\pm 8$V and impose an external electric field in the $x$ direction corresponding to a 500V potential drop across the domain. The resulting shear rate from the electroosmotic flow is about $0.4~\mathrm{ns}^{-1}$.

\begin{figure}
  \centering
  \includegraphics[width=.24\textwidth]{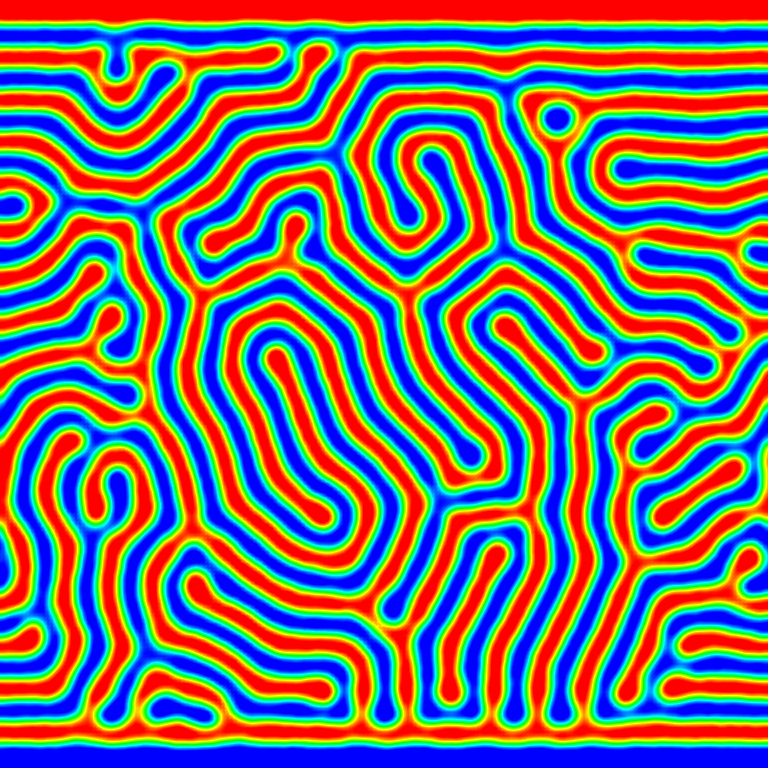}
  \includegraphics[width=.24\textwidth]{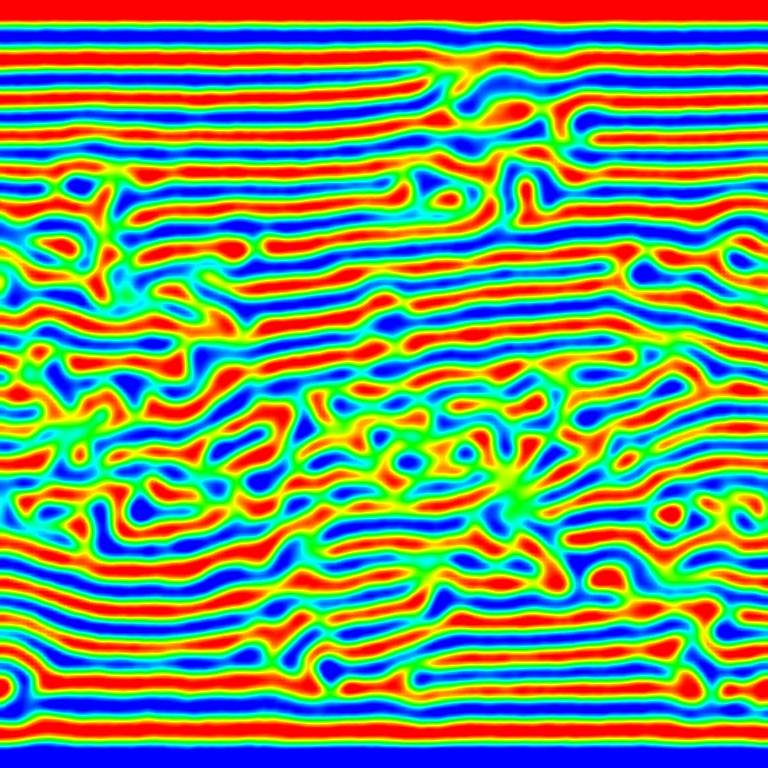}
 % 
 % \vspace{.05in}
  \includegraphics[width=.24\textwidth]{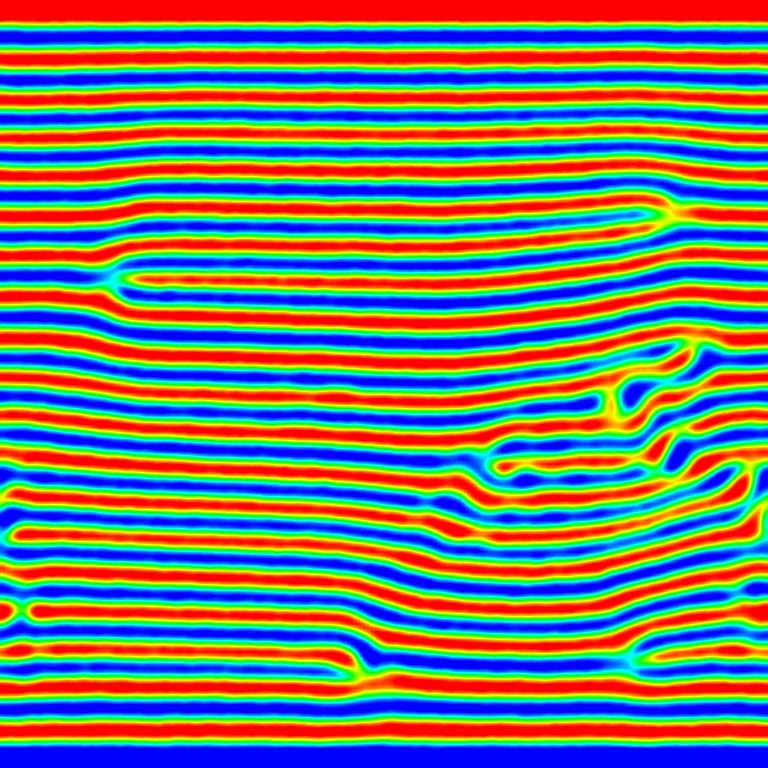}
  \includegraphics[width=.24\textwidth]{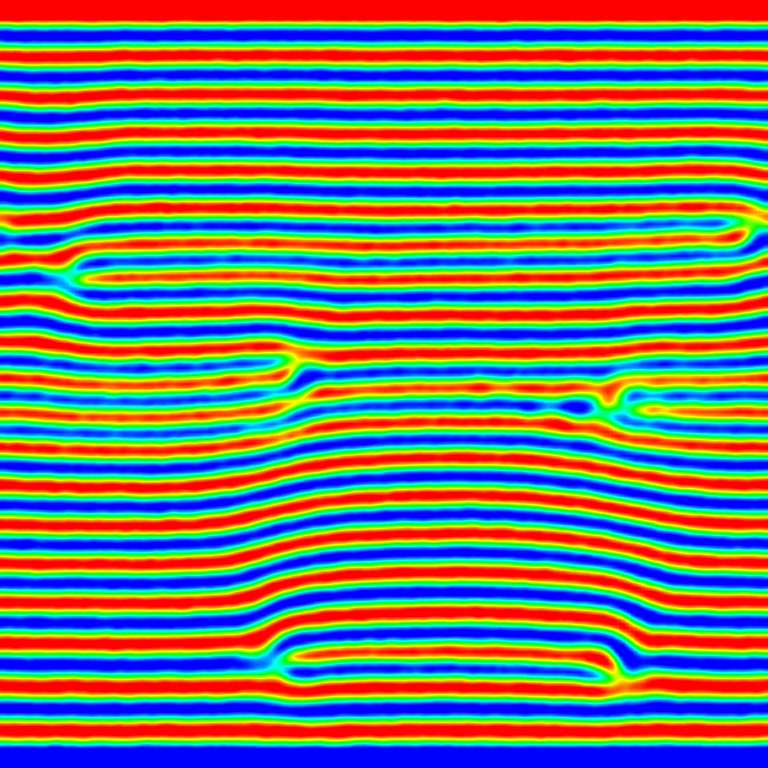}
  \caption{Time evolution of the cation concentration in the RTIL under electroosmotic shear; images are at $t=0, 2,  5$, and $20$~ns.}
\label{fig:2D_shear}
\end{figure}
The temporal evolution of the resulting sheared system
is shown in Figure \ref{fig:2D_shear}.  For the strong imposed electric field considered in this example, the shearing first destroys the serpentine patterns and then structure is restored, developing a striated pattern parallel to the imposed electric field with a few long-lived defects.
Molecular dynamics studies also indicate that nanostructures in RTILs persist in the presence of a strong shear.\cite{Raju2008,Butler2014}

\section{Conclusions}\label{sec:conclusions}

The computational model presented here is the first step towards a mesoscale simulation capability for room temperature ionic liquids that includes both hydrodynamics and thermal fluctuations. 
The fluctuating hydrodynamics formulation for RTILs is a useful intermediary, mesoscale theory bridging microscopic models, such as molecular dynamics, and macroscopic models.%\MarginPar{given our reliance on Onsager theory \& stochastic fluxes, I would argue we developed a phenomenological macroscopic model. Maybe we can repeat what was said in the intro ie ``coarse-grained lattice models"? -SC}
The low Mach number formulation avoids the severe time step restriction associated with previous compressible formulations. We
have demonstrated in both two and three dimensions that resulting methodology reproduces the microscopic structuring observed in RTILs.
We also used the methodology to show that the differential capacitance decreases with applied voltage which is a characteristic of ionic liquids.
The morphology patterns observed in this capacitance geometry were significantly different depending on whether the simulations included or excluded thermal fluctuations.
Finally, the shear electroosmosis example shows that the methodology can be used to model electrokinetic flows.

The present model can be enhanced and extended in several important directions in the future. In this paper we considered a two-component ionic fluid with symmetric ions; however, the RTIL model can be generalized to arbitrary multi-component mixtures. 
(See \citenum{peraud2016} for a more general FHD model of
multicomponent electrolyte solutions.)
This will allow us to consider RTILs composed of dissimilar ions as well as investigate the effect of a polar solvent on the structural, thermodynamic, and electrical properties.

As is commonly assumed in RTIL modeling we assumed the permittivity to be a constant. A more realistic version of the model makes the permittivity a function of concentration, which is important for the study of RTIL mixtures. The implementation requires modifying the calculation of the Poisson equation and the Maxwell stress tensor; a greater challenge is determining an accurate functional form of $\epsilon(c)$. A related extension is to include dielectric relaxation~\cite{Stoppa2010} by treating the local permittivity (or equivalently, the local polarization density) as a stochastic quantity whose dynamics are given by a Langevin equation. 

The increased time step associated with the low Mach number fluctuating hydrodynamics model makes it possible to investigate long time dynamics and three dimensional effects.
Many ionic liquids tend to have glassy behaviors~\cite{sangoro2012charge, frenzel2019charge} that make the equilibration of molecular dynamics simulations particularly challenging. Our fluctuating hydrodynamics model permits numerical explorations of RTIL regimes with slow dynamics.
The capability to perform three dimensional simulations is particularly important for future investigations of the structure and dynamics of the double layer. The transitions between lateral arrangements of counter and co-ions at electrified interfaces greatly affects the capacitance and dynamics of the double layers, and this is related to the three-dimensional nature of the double layer in ionic liquids~\cite{kornyshev2014three}. As part of this type of investigation, more physically realistic boundary conditions that can capture wetting effects at electrode-RTIL interfaces will need to be introduced.

Generalization of the methodology to more complex boundary conditions and geometries would allow us to explore capacitance enhancement in nanopores~\cite{pean2014dynamics}, which are important in the development of supercapacitors based on nanostructured electrodes. The treatment
%\MarginPar{changho's stuff??? -- JBB}
of electrochemical effects at electrode boundaries would be a topic for future work.

Finally, the FHD methodology presented here lays the foundation for hybrid, or ``heterogeneous" methods \cite{Garcia_99,abdulle:2012,Hybrid2020} that couple a continuum hydrodynamic description to a more detailed microscopic model, e.g., molecular dynamics.
For these types of ``Adaptive Algorithm'' hybrids, the continuum model needs to include thermal fluctuations in order to correctly capture the behavior in the region being
modeled with the microscopic model \cite{hybrid:2010}.
This type of hybrid model would enable simulations to use a microscopic representation locally where molecular-level accuracy is desired, such as near electrode surfaces, while using a less expensive continuum-based model in the remainder of the domain.

\section*{Acknowledgements}

This work was supported by the U.S.~Department of Energy, Office of Science, Office of Advanced Scientific Computing Research, Applied Mathematics Program under contract No.~DE-AC02-05CH11231.
This research used resources of the National Energy Research Scientific Computing Center, a DOE Office of Science User Facility supported by the Office of Science of the U.S. Department of Energy under Contract No.~DE-AC02-05CH11231.%\MarginPar{Please check. AG}

%%%%%%%%%%%%%%%%%%%%%%%%%%%%%%%%%%%%%%%
%%%% Appendix
%%%%
\appendix
\section{Derivation of Inviscid Hydrodynamic Equation}\label{section:EL_appendix}
Here we derive the inviscid form of the momentum equation \eqref{eq:low_mach_eqs}
using the calculus of variations. It will be useful below to introduce the notation
\begin{equation}\label{eq:adv_deriv}
    \Dt f := f_t + u\cdot \nabla f
\end{equation}
for the advective derivative of scalar $f$ with respect to the velocity field $u$. For the case when $f$ is a vector of functions, equation \eqref{eq:adv_deriv} is understood component-wise, so that in Einstein notation
\begin{equation}
    \left(\Dt f\right)_i = \left(f_i\right)_t + u_j \pderiv{f_i}{x_j}
\end{equation}

Starting from the action 
\begin{equation}
    S = \int dt \int d\mathbf{r} \, \frac12 \rho_0 |u|^2 - \left( \rho_0 g(c,\nabla c) + \rho_0 z(2c-1)\phi - \frac12 \epsilon |\nabla \phi|^2\right),
\end{equation}
which is simply the space-time integral of the kinetic energy of the fluid minus the time integral of the free energy functional $G$ (see eqn.~\eqref{eq:FE}), 
we add the constraints that the flow $u$ is divergence free and that the concentration $c$ is advected by the flow. The action then becomes 
\begin{align}
    S = \int dt \int d\mathbf{r} \, \frac12 \rho_0 |u|^2 &- \left( \rho_0 g(c,\nabla c) + \rho_0 z(2c-1)\phi - \frac12 \epsilon |\nabla \phi|^2\right)  \nonumber \\
    &+ \rho_0 \chi \left(\nabla \cdot u \right) +  \rho_0 \lambda \left(c_t + \nabla \cdot (uc)\right), 
\end{align}
where $\chi$ and $\lambda$ are the Lagrange
multipliers. 
These extra constraints are necessary for nontrivial dynamics for the velocity field; enforcing that the action $S$ is stationary with respect to variations in $u$ results in 
\begin{equation}\label{eq:clebsch}
    u = c\nabla \lambda + \nabla \chi
\end{equation}
after integrating by parts and assuming the space of trial functions is such that the boundary terms vanish. This is the
well-known Clebsch representation \cite{Clebsch_1859,salmon1988hamiltonian}. 
Variations with respect to $\phi$ and $\chi$ result in the Poisson equation of electrostatics and the divergence-free constraint
\begin{align}
    -\epsilon \nabla^2 \phi &= \rho_0 z(2c-1) = q^f \\
    \nabla \cdot u &= 0. \label{eq:div_free_constraint}
\end{align}
Variations with respect to $\lambda$ and using \eqref{eq:div_free_constraint} result in the constraint
\begin{equation}
    \Dt c = 0. 
\end{equation}
Consider now the advective derivative of the Clebsch representation \eqref{eq:clebsch}, and note that $\Dt$ is a linear operator that obeys the standard product rule of differential calculus $\Dt \left(fg\right) = g \Dt f + f \Dt g$, as well as the commutation relation
\begin{equation}\label{eq:comm_rule}
    \Dt \left(\nabla f \right) = \nabla \left(\Dt f\right) - \nabla u \nabla f,
\end{equation}
where the second term is a matrix-vector product and $(\nabla u)_{ij}$ is defined as $\partial u_i/\partial x_j$. 
Hence
\begin{flalign}
    && \Dt u &= \Dt \left(c \nabla \lambda\right) + \Dt\left( \nabla \chi\right) &&\\
    &&   &= c \Dt \left(\nabla \lambda \right) + \left(\Dt c\right) \nabla \lambda + \Dt \left( \nabla \chi\right),   && (\text{Eq. } \eqref{eq:comm_rule}) \\
    &&  &= c \nabla \left(\Dt \lambda\right) - c\nabla u \nabla \lambda + \nabla\left(\Dt \chi\right) - \nabla u \nabla\chi && \left(\Dt c=0\right) \\
    && &= c \nabla \left(\Dt \lambda\right) + \nabla\left(\Dt \chi - |u|^2/2\right) && \left(u = c\nabla\lambda + \nabla\chi\right). \label{eq:rhs_Dt_u}
\end{flalign}
The term $c\nabla\left(\Dt \lambda\right)$ will be the source of the Lorentz force density \eqref{eq:lorentz_force} and the divergence of the interfacial reversible stress \eqref{reversible_stress}.

It remains to consider variations of the action $S$ with respect to the concentration $c$. Grouping together the contributions to the specific free energy modeling the enthalpy and entropy of mixing into a single term
\begin{equation}
    g^{\rm loc}(c) := \frac{k_B T}{m}\left(g^{\rm id}(c) + g^{\rm ex}(c)\right),
\end{equation}
eqn.~\eqref{eq:g_no_electrostatic} becomes
\begin{equation}
    g(c,\nabla c) = g^{\rm loc}(c) + \frac{k_B T}{m} \frac12 \kappa \left|\nabla c\right|^2.
\end{equation}
% Furthermore, it is advantageous to integrate the contribution $\lambda \rho_0 \Dt c$ to $S$ by parts; assuming that the space of trial functions is such that the boundary terms vanish, the action becomes
% \begin{align}
%     S = \int dt \int d\mathbf{r} \, \frac12 \rho_0 |u|^2 &- \left( \rho_0 g^{\rm loc}(c) + \rho_0 \frac{k_B T}{m} \frac12 \kappa |\nabla c|^2 + \rho_0 z(2c-1)\phi - \frac12 \epsilon |\nabla \phi|^2\right) \\
%     &+ \chi \left(\nabla \cdot u \right) - \rho_0 c \Dt \lambda , 
% \end{align}
% where the divergence-free condition was also used. 
Variations with respect to $c$ then result in 
\begin{align}
    \Dt \lambda &= -\pderiv{g^{\rm loc}}{c} + \frac{k_B T}{m} \kappa \nabla^2 c - 2 z \phi \\
    \implies c\nabla\left(\Dt \lambda\right) &= -c\nabla \left(\pderiv{g^{\rm loc}}{c}\right) + \frac{k_B T}{m} \kappa c \nabla\left( \nabla^2 c\right) - 2zc\nabla \phi \label{eq:pre_Dt_lambda}
\end{align}
After manipulating the three terms on the right hand side of \eqref{eq:pre_Dt_lambda}, we will insert them into \eqref{eq:rhs_Dt_u}. The third term can be written as 
\begin{align}
    -2zc\nabla \phi &= -z(2c-1)\nabla\phi - z\nabla\phi \\
    &= -\left(\frac{1}{\rho_0}\right) q^f \nabla \phi - z\nabla \phi \\
    &= \frac{1}{\rho_0} \nabla \cdot \MaxwellTensor - z\nabla \phi \label{eq:third_term}
\end{align}
using the definition of the Maxwell stress tensor \eqref{eq:lorentz_force}. The second term can be written by combining the two product rule identities:
\begin{equation}
    c \nabla\left(\nabla^2 c\right) = \nabla\left(c \nabla^2 c\right) -  \nabla^2 c \, \nabla c
\end{equation}
and 
\begin{align}
    \nabla^2 c \, \nabla c  &= \nabla \cdot \left(\nabla c \otimes \nabla c\right) - \frac12 \nabla\left(\left|\nabla c\right|^2 \right) \\
    &= \nabla \cdot \left(\nabla c \otimes \nabla c - \frac12 \left|\nabla c\right|^2 \mathbb{I}\right)
\end{align}
so that 
\begin{align}
    \frac{k_B T}{m} \kappa c \nabla\left( \nabla^2 c\right) &= \nabla \left(\frac{k_B T}{m} \kappa c \nabla^2 c\right) -  \frac{k_B T}{m} \kappa \nabla \cdot \left(\nabla c \otimes \nabla c - \frac12 \left|\nabla c\right|^2 \mathbb{I}\right) \label{eq:rev_stress_identity1}\\
    &= \nabla \left(\frac{k_B T}{m} \kappa c \nabla^2 c\right) + \frac{1}{\rho_0}\nabla \cdot \left(\frac{\rho_0 k_B T\kappa}{m} \left[\frac12 \left|\nabla c\right|^2 - \nabla c \otimes \nabla c\right]\right) \\
    &= \nabla \left(\frac{k_B T}{m} \kappa c \nabla^2 c\right) + \frac{1}{\rho_0}\nabla \cdot \ReversibleStress \label{eq:second_term}
\end{align}
by definition of the interfacial reversible stress tensor \eqref{reversible_stress}. Lastly, the first term on the right hand side of equation \eqref{eq:pre_Dt_lambda} can be re-written using the Gibbs-Duhem relation of thermodynamics \cite{DM_63}, which says for our isothermal, two-component mixture
\begin{equation}
    \nabla p = \rho_1 \nabla \mu_1 + \rho_2 \nabla \mu_2  
\end{equation}
where $p$ is the thermodynamic pressure and $\mu_k = \partial g^{\rm loc}/\partial c_k$. 
Since $\rho_k = \rho_0 c_k$, $c=c_1$, and $c_2 = 1-c_1$, 
\begin{equation}
    \pderiv{g^{\rm loc}}{c_2} = - \pderiv{g^{\rm loc}}{c_1},
\end{equation}
and hence the Gibbs-Duhem relation reduces to
\begin{align}
    \nabla p &= \rho_0(2c-1)\nabla \left(\pderiv{g^{\rm loc}}{c}\right) \\
    \implies -c\nabla\left(\pderiv{g^{\rm loc}}{c}\right) &= -\frac12 \frac{\nabla p}{\rho_0} - \frac12 \nabla \left(\pderiv{g^{\rm loc}}{c}\right). \label{eq:gibbs_duhem}
\end{align}
Inserting the relations \eqref{eq:gibbs_duhem}, \eqref{eq:second_term}, and \eqref{eq:third_term} into the $c\nabla (\Dt \lambda)$ term in \eqref{eq:rhs_Dt_u} then results in
\begin{align}
    \Dt u &= c \nabla \left(\Dt \lambda\right) + \nabla\left(\Dt \chi - |u|^2/2\right) \\
    &=  \frac{1}{\rho_0}\nabla \cdot \ReversibleStress  + \frac{1}{\rho_0}\nabla \cdot \MaxwellTensor + \nabla\left(\Dt \chi + \frac{k_B T \kappa}{m} c \nabla^2 c - \frac12 |u|^2  - \frac12 \frac{1}{\rho_0}p - \frac12 \pderiv{g^{\rm loc}}{c}-z\phi\right),
\end{align}
and after identifying 
\begin{equation}
    \pi := -\left(\rho_0\Dt \chi + \frac{\rho_0 k_B T \kappa}{m} c \nabla^2 c - \frac12 \rho_0 |u|^2  - \frac12 p - \frac12 \rho_0 \pderiv{g^{\rm loc}}{c}-\rho_0 z\phi\right)
\end{equation}
as a perturbational pressure, we arrive at the inviscid form of the momentum equation in \eqref{eq:FE}
\begin{equation}
    (\rho_0 u)_t + \nabla \cdot \left(\rho_0 uu\right) + \nabla \pi = \nabla \cdot \MaxwellTensor + \nabla \cdot \ReversibleStress
\end{equation}
as desired. $\square$

\bibliography{rtil}

\end{document}